\documentclass{aastex631}  

\usepackage{hyperref}  
\usepackage{booktabs}  
\usepackage{amsmath}   
\usepackage{natbib}    
\usepackage{url}       
\usepackage{footnote}  
\usepackage{graphicx}  
\usepackage{capt-of} 
\usepackage{float}     

\usepackage{amssymb}   
\usepackage{epstopdf}  

\begin{document}

\title{Period Analysis of Eclipsing Cataclysmic Variable Stars} 
\author{Mennatalla Mahmoud Ellaqany, Valeria Garcia-Lopez, Emily S. Hatten, Mridul Agarwal, David A. Moffett}
\affiliation{Department of Physics\\ Furman University\\ Greenville, South Carolina 29613, USA}
\date{\today}
\keywords{Eclipsing Binary Stars --- 
Cataclysmic Variable Stars --- Period Determination --- Orbital Evolution}

\section*{Abstract}
We have performed a study of the orbital properties of seven eclipsing cataclysmic variable (CV) binary systems by analyzing photometric time series from the Transiting Exoplanet Survey Satellite (TESS). We employed Python code to determine the eclipse epochs and orbital periods for each system, and constructed $O-C$ diagrams from observed and predicted eclipse epochs.  By analyzing the $O-C$ diagrams of our target CVs, we have constrained values for changes in orbital period with time. Our targets include a sample of sources from each class of non--magnetic, eclipsing CVs: dwarf novae variables, Z Cam type, and U Gem subclasses.  We include in our study classical novae variables, nova--like variables (including the VY Scl and UX UMa subclasses), and recurrent novae variable stars.  We approached this project with goals of developing time series analysis techniques for future undergraduate-level studies of eclipsing CVs, and how they may contribute to the understanding of their orbital evolution. 

\section{Introduction}
Eclipsing Cataclysmic Variables (ECVs) are a subset of cataclysmic variable (CV) systems whose high inclination angles give us an unambiguous measurement of orbital period 
and eclipse phase of the white dwarf and companion star.  Observations of ECV light curves allow observers to study the orbital dynamics of the binary system and the accretion disk.  Eclipsing CVs sample several classes of cataclysmic variables, and can be divided into even more subclasses which have their own distinct accretion mechanisms and outburst behaviors, although we will only be focusing on dwarf novae (DN) and classical novae (CN). Dwarf Novae systems exhibit periodic outbursts, driven by the gravitational energy released from an instability in the binary's accretion disk or from a sudden variations in the mass-transfer process.  The outburst intervals range from days to decades, and the outburst period can last from two to 20 days. Dwarf nova systems are divided further into three subclasses: U Gem, Z Cam and SU UMa. Classical Novae are a subclass of variable novae systems, which have a high and steady accretion rate but do not exhibit large-scale outbursts like Dwarf Novae. The typical brightness range for CN range from magnitude 6 to greater than 19. Brighter novae have shorter durations than fainter novae. CN outbursts are usually modeled as thermonuclear runaways originating in accreted hydrogen-rich material on the surface of the primary star.
\\
Studying different types and subtypes of eclipsing cataclysmic variable systems allows us to break down the accretion properties while under different conditions. Understanding the accretion process and properties contribute to understanding the long term evolution process of ECVs. We hope to observe these systems using TESS archival data to model the orbital period and period derivative (if present) of CV systems, with the intent to search for other periodic and aperiodic behavior associated with the accretion disk, hotspots, and any other unknown mechanisms that might impact the light curve. 

\clearpage
\section{Observation Data}
Our small survey consists of the light curves of eclipsing cataclysmic variable (CV) stars observed by TESS, initially processed at NASA's Science Processing Operations Center (SPOC). TESS was designed to discover exoplanets that transit stars bright enough to enable follow-up spectroscopic observations for determining planet masses and atmospheric compositions. Since its launch in 2018, TESS has observed over 200,000 main-sequence dwarf stars using four wide-field optical cameras, detecting periodic brightness variations caused by planets passing in front of their host stars. The satellite's unique elliptical high Earth orbit provides a stable platform for continuous photometric measurements, surveying over $85\%$ of the sky and focusing on stars brighter and closer than those observed by the Kepler mission \citep{barclay_tess}.  The SPOC pipeline identifies potential planetary transits, corrects systematic errors, and conducts diagnostic tests, distinguishing TESS's data processing from that of Kepler \citep{barclay_tess}. In this study, we are evaluating the target light curve files, which include photometric analysis and systematically corrected time series data, to investigate the period behavior of eclipsing CVs. 
\\
We acquired 2-minute cadence photometric data of normalized flux versus BTJD (Barycentric TESS Julian Date) of seven CVs: QZ Aur, EX Dra, EX Hya, AY Psc, DO Leo, GY Cnc, and HBHA 4204--09. The data comprises of multiple segments extending over BTJD ranges that together span about 3 to 4 years, providing a long dataset suitable for our study. This duration is sufficient for observing many eclipses, as most CVs have an orbital period of a few hours. We retrieved the data using a Python code that accesses MAST via the unique TIC (TESS Input Catalog) ID number assigned to each object. Our objective is to construct an $(O-C)$ diagram and forecast their rate of period change, denoted as $\dot{P}$. The observational details of CVs we chose in our study can be found in Table \ref{tab:tess_data}.  

\begin{table}[h!]
    \centering
    \setlength{\tabcolsep}{15pt}
    \begin{tabular}{lccc}
        \hline
        \hline
        \textbf{TIC} & \textbf{CV} & \textbf{Class} & $\mathbf{\triangle Y}$ \textbf{(years)}\\
        \hline
        \hline
        3034524   & QZ Aur        & C Nova & 2019-2023 \\
        219107776 & EX Dra        & D Nova & 2019-2024 \\
        9560142 & EX Hya        & D Nova & 2019-2023 \\
        346897118 & AY Psc        & D Nova & 2021-2023 \\
        61285257  & DO Leo        & D Nova & 2020-2023 \\
        238609772 & GY Cnc        & D Nova & 2021-2023 \\
        304628774 & HBHA 4204--09 & D Nova & 2019-2024 \\
        \hline
        \hline
    \end{tabular}
    \caption{TIC identification, source name, CV type, and observed time range of our CV sample.}
    \label{tab:tess_data}
\end{table}
\raggedbottom
\section{Data Processing Methods}
\label{sec:data-processing}
\subsection{Data Smoothing and Interpolation}
After importing the data from MAST, the light curves then undergo a few data processing steps. First, we remove any $NaN$ values, and remove any bright, long-term features in the photometric flux data ($i.e.$, outbursts) using a Savitzky-Golay filter. The way the Savitzky-Golay filter smoothes light curves is through convolution. We used a sliding window (a number of consecutive data points) and then fitted a second degree polynomial to the flux values to get a smoothed value for each data point. The coefficients of the polynomial are obtained using least squares regression. After computing a smoothed flux value, it is subtracted from the original flux value to get a 'detrended' residual light curve.  Large scale flux variations are removed, leaving smaller variations like the eclipses unchanged. The Savitzky-Golay filter method is also used in the same way to smooth the eclipses in the light curves by fitting a Gaussian to a smaller sliding window instead. We apply a spline to the residual data to create a $12$-second cadence light curve while skipping over observing gaps to avoid introducing unwanted artifacts. We use cubic splines to fill in the gaps between $2$-minute cadence data points while skipping over gaps of $20$ minutes or longer. An example of a detrended light curve for EX Dra can be found in Figure \ref{fig:lightcurve_exdra} below.  The light curves of the other six sources can be found in Figure \ref{fig:light_curves} in the Appendix.

\begin{figure}[H]
    \centering
    \includegraphics[width=0.7\linewidth]{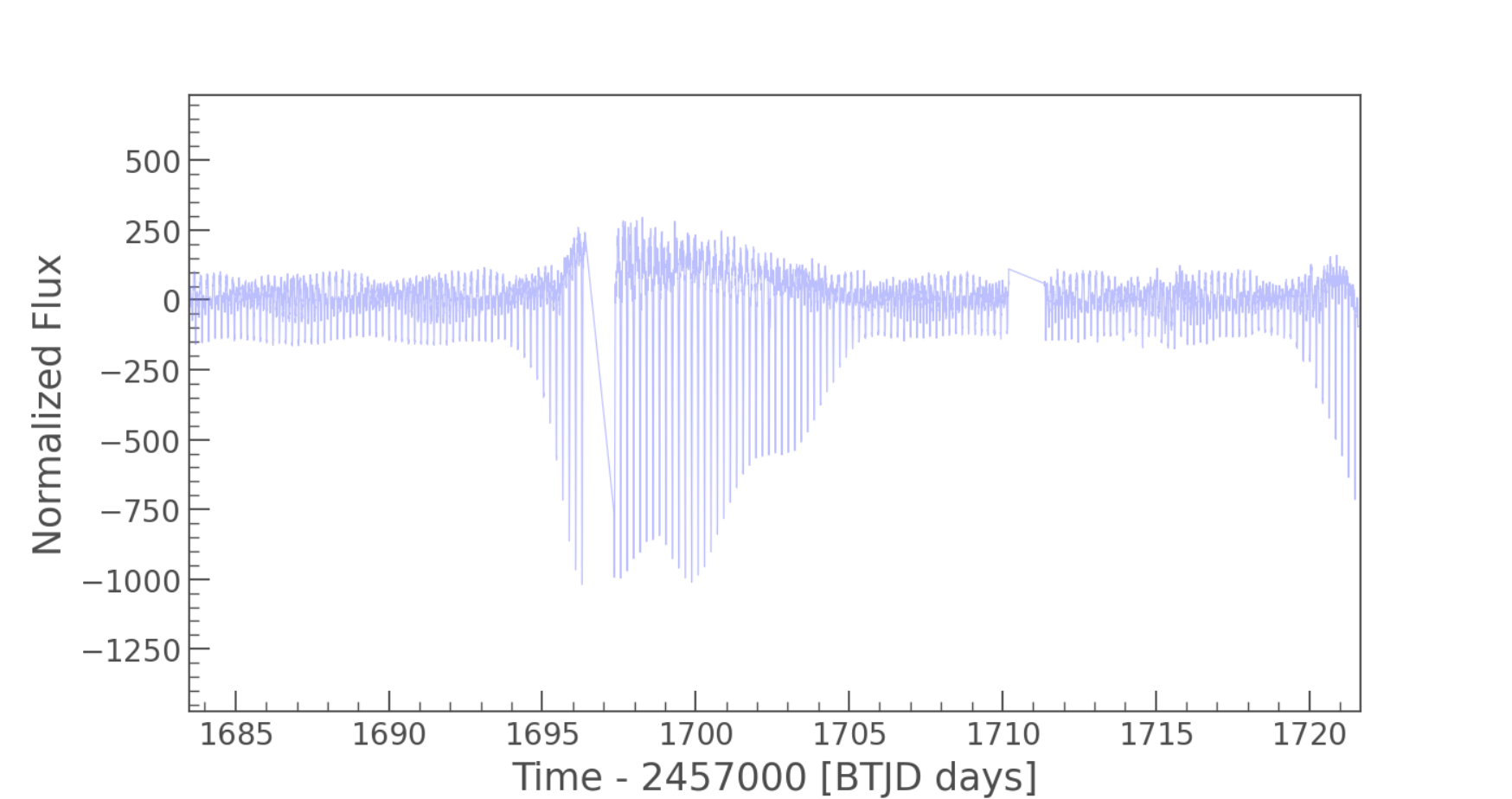}
    \caption{Detrended TESS light curve of Ex Dra. Time is in units of Barycentric Julian Day (BTJD), and amplitude is in units of normalized flux.}
    \label{fig:lightcurve_exdra}
\end{figure}

\subsection{Periodogram Analysis}
\begin{figure*}[h!] 
    \centering
    \includegraphics[width=0.65\linewidth]{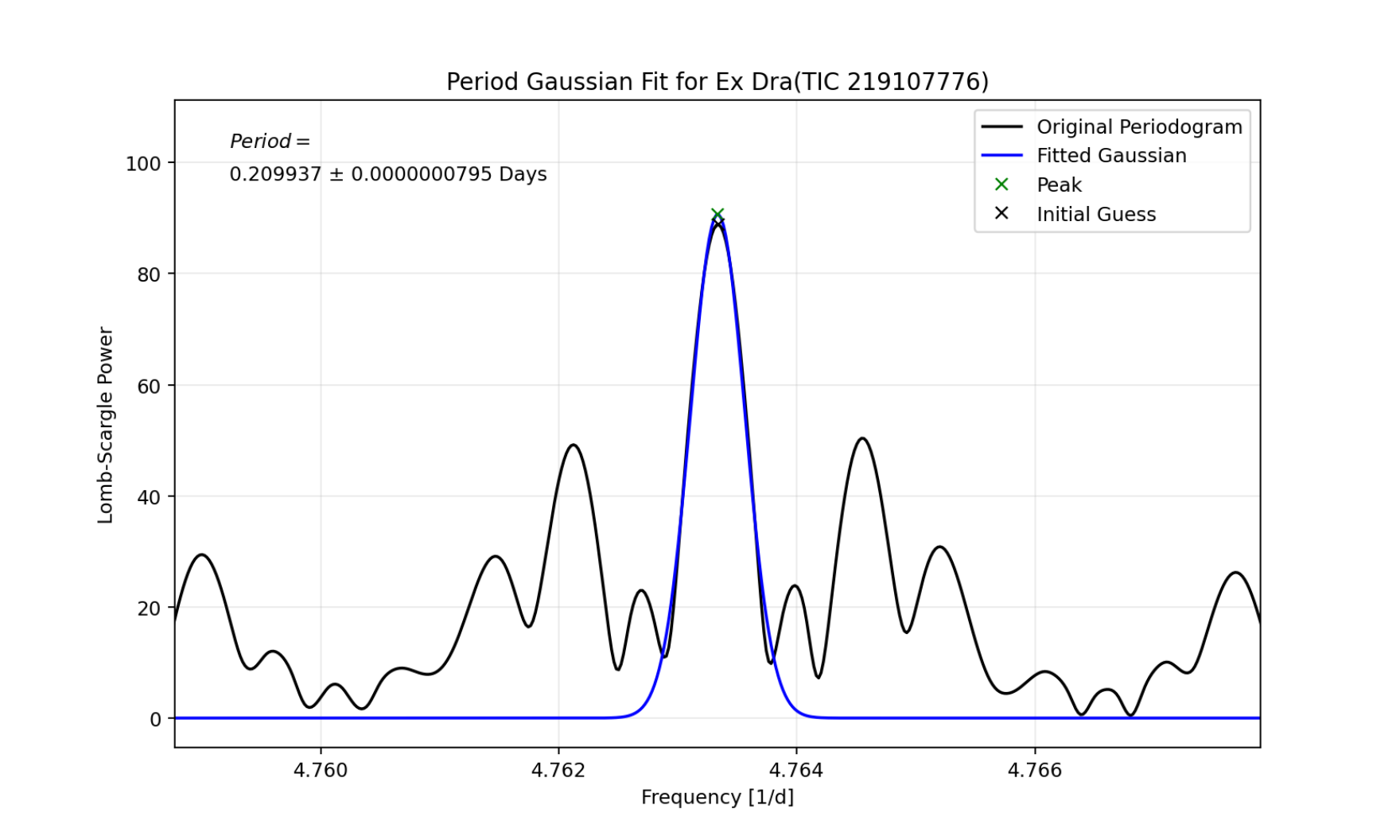}  
    \caption{Lomb-Scargle periodogram for EX Dra. The highest peak was fitted by a Gaussian function for the orbital period.}
    \label{fig:periodogram_exdra}
\end{figure*}

\begin{figure*}[h!]
    \centering

    \begin{minipage}{0.45\textwidth}
        \centering
        \includegraphics[width=\linewidth]{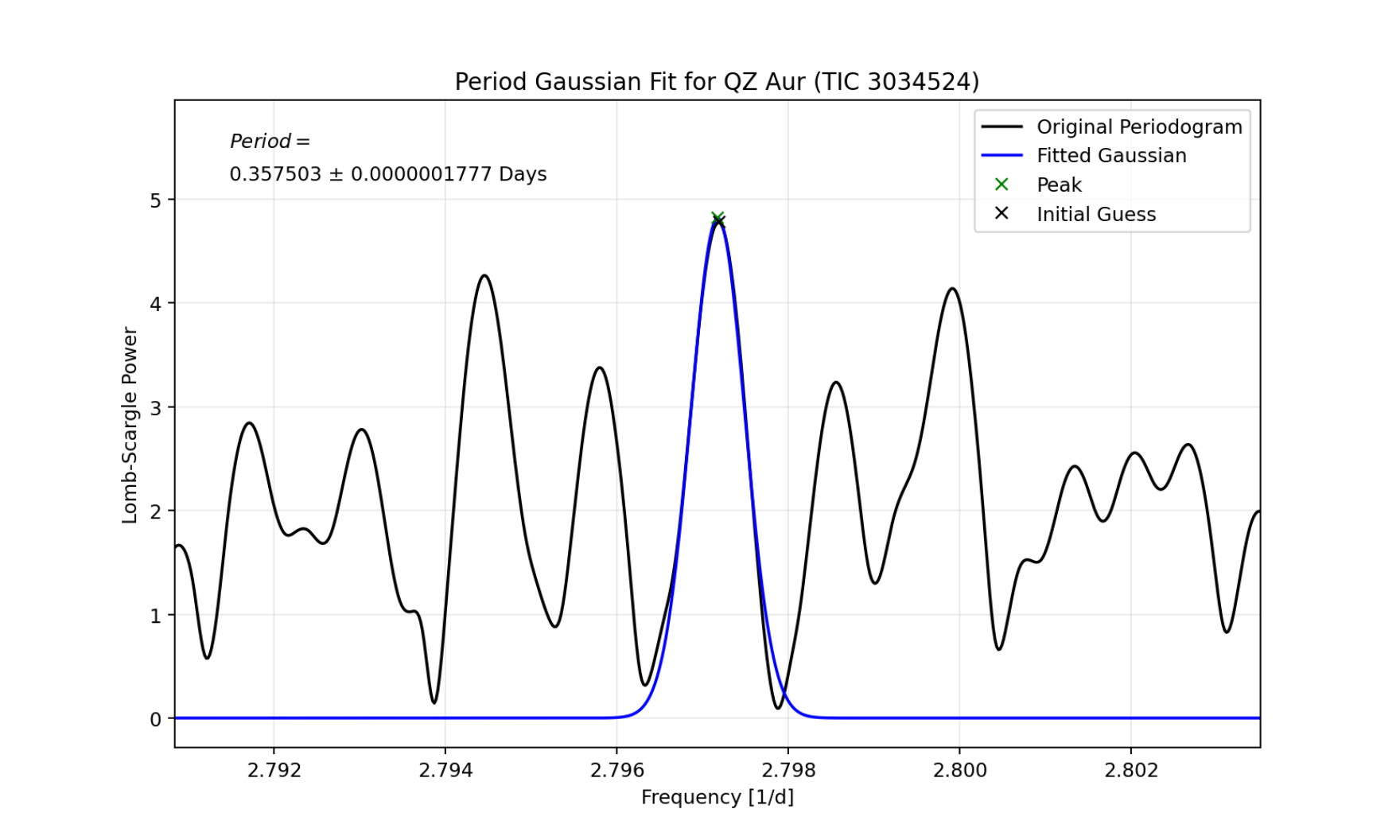}
        \caption{QZ Aur}
    \end{minipage}%
    \hfill
    \begin{minipage}{0.45\textwidth}
        \centering
        \includegraphics[width=\linewidth]{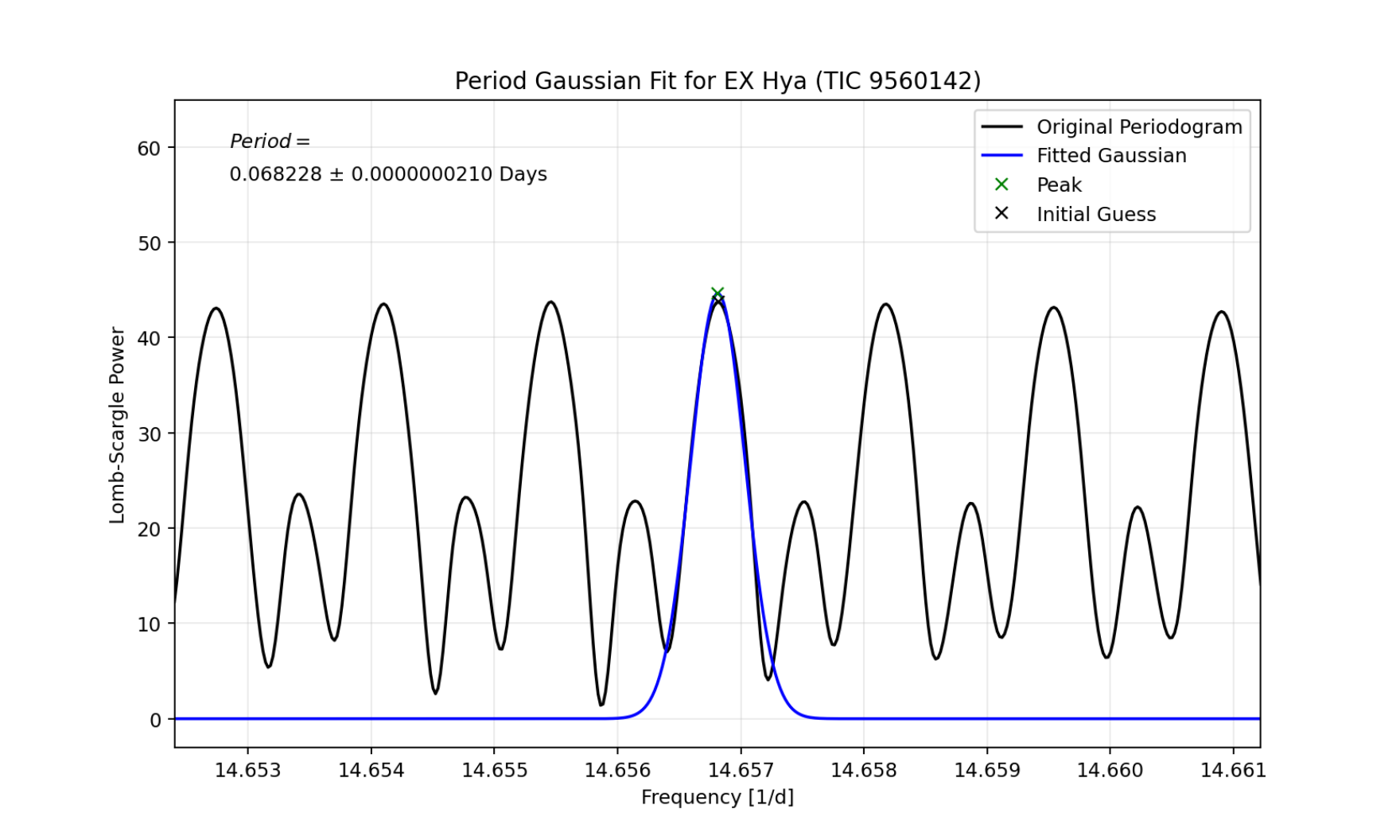}
        \caption{EX Hya}
    \end{minipage}

    \vspace{1em} 

    \begin{minipage}{0.45\textwidth}
        \centering
        \includegraphics[width=\linewidth]{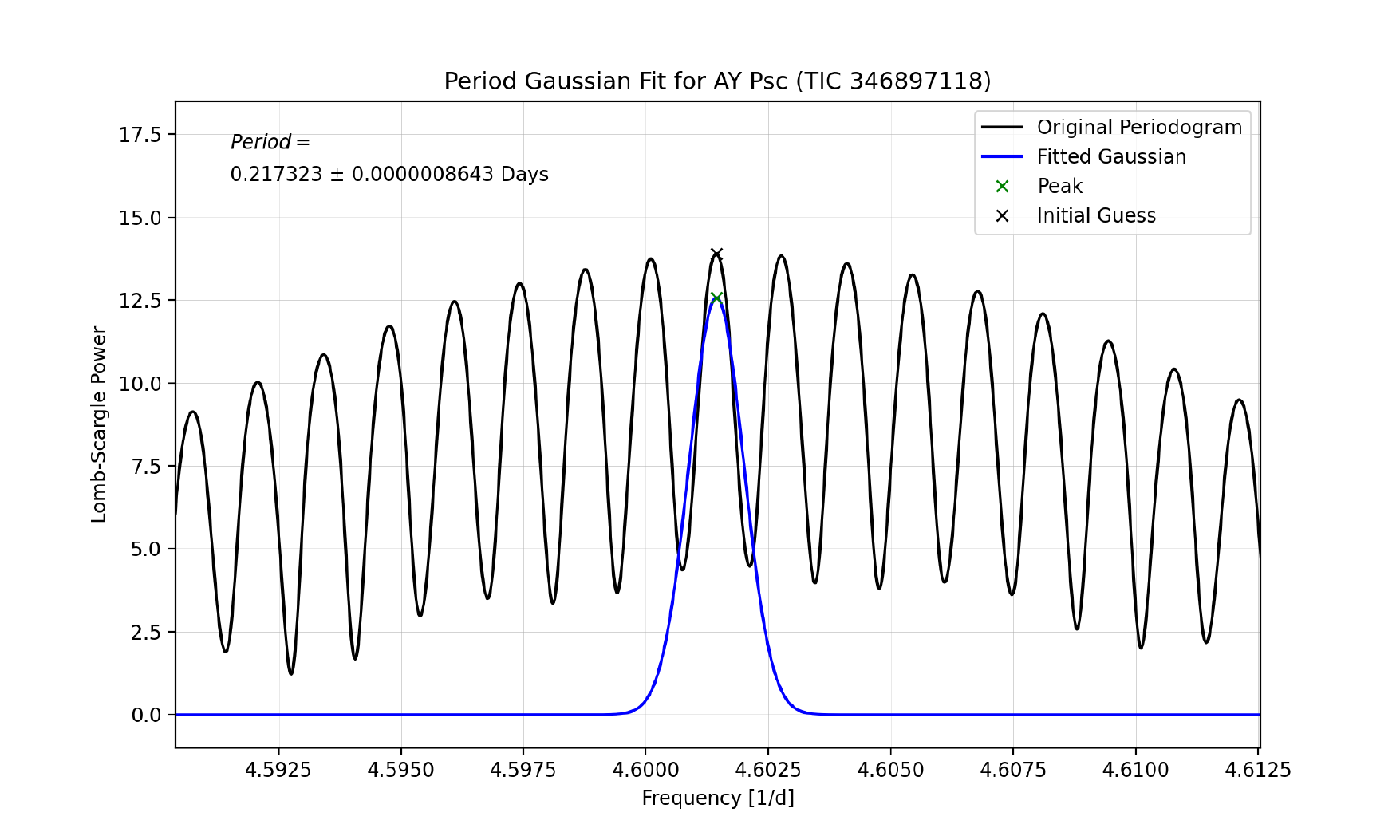}
        \caption{AY Psc}
    \end{minipage}%
    \hfill
    \begin{minipage}{0.45\textwidth}
        \centering
        \includegraphics[width=\linewidth]{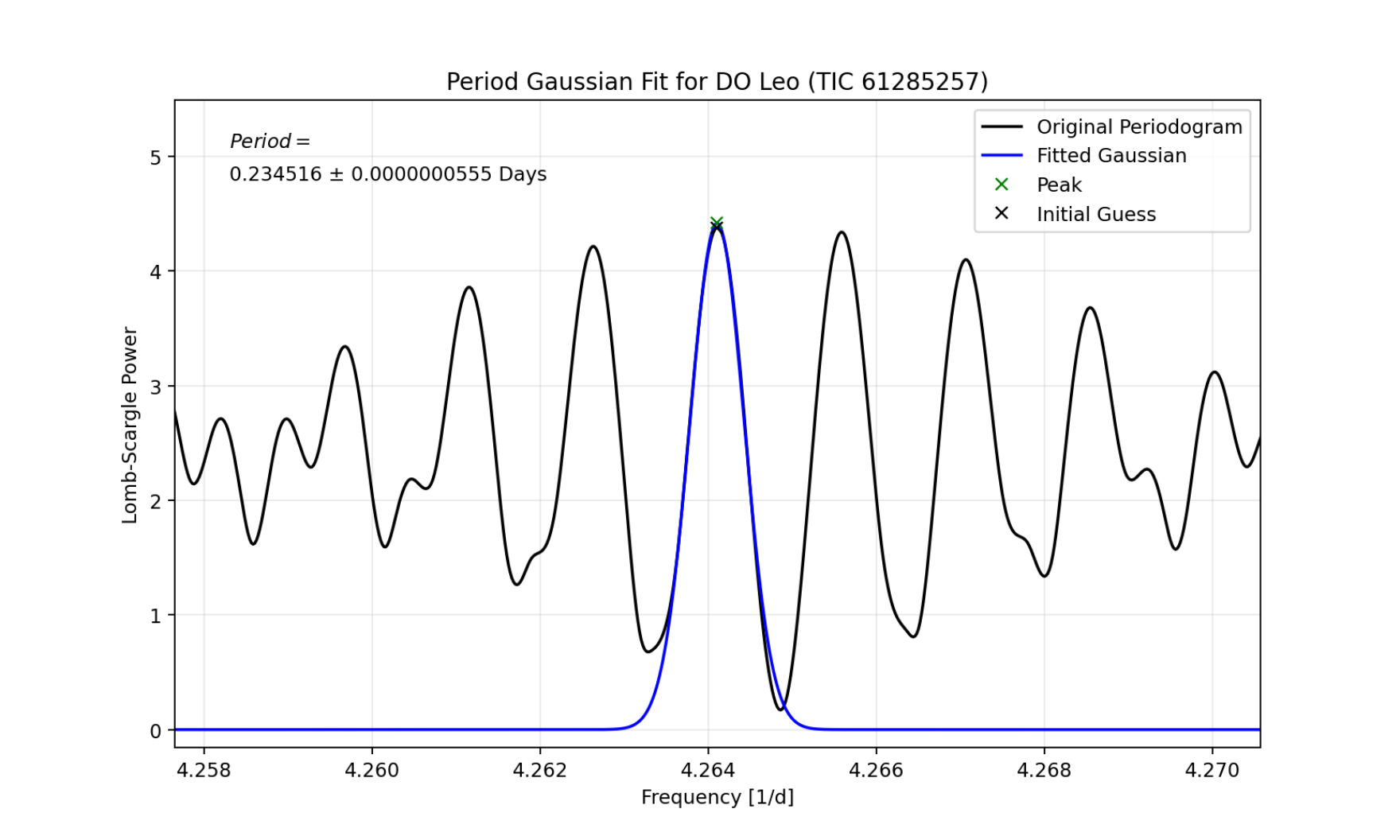}
        \caption{DO Leo}
    \end{minipage}

    \vspace{1em} 

    \begin{minipage}{0.45\textwidth}
        \centering
        \includegraphics[width=\linewidth]{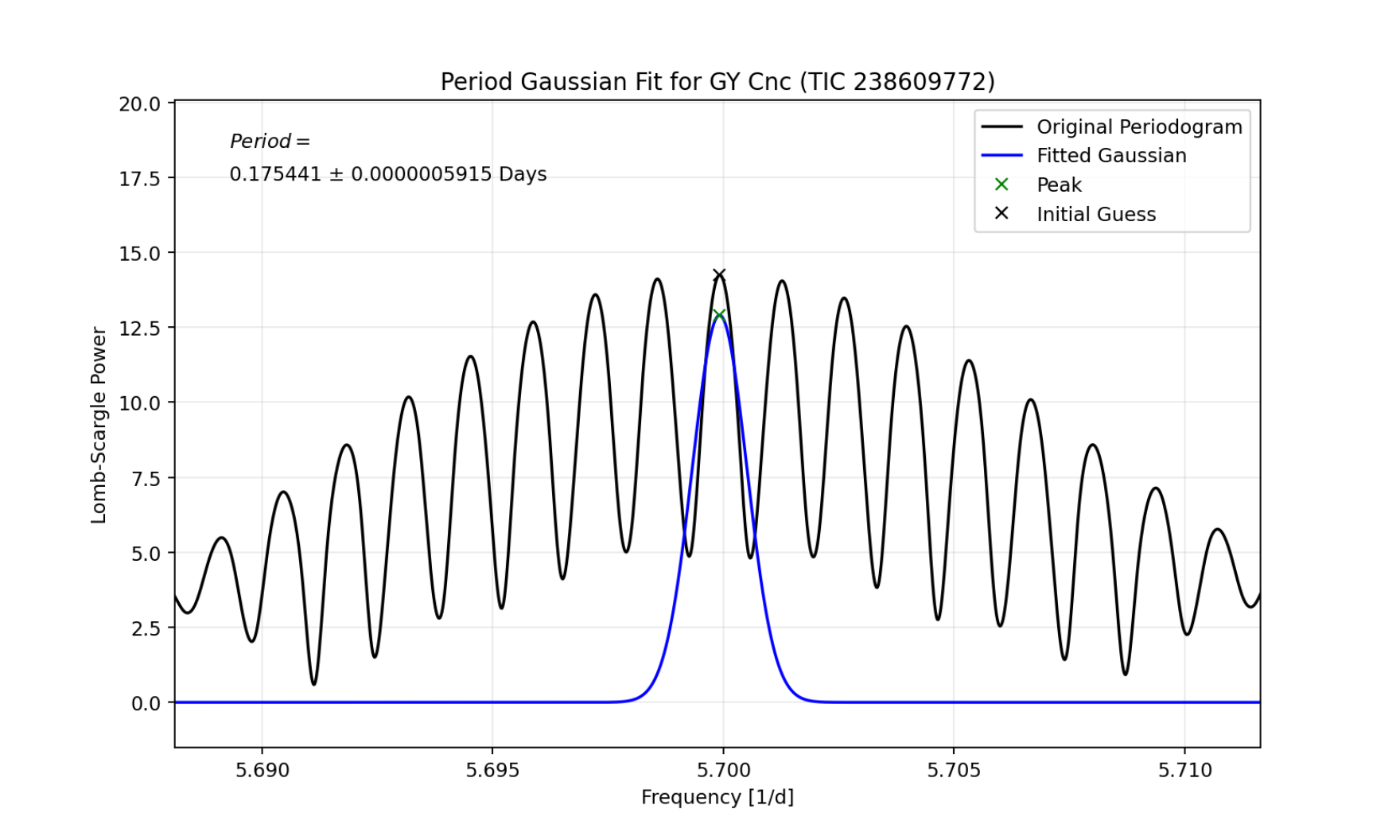}
        \caption{GY Cnc}
    \end{minipage}%
    \hfill
    \begin{minipage}{0.45\textwidth}
        \centering
        \includegraphics[width=\linewidth]{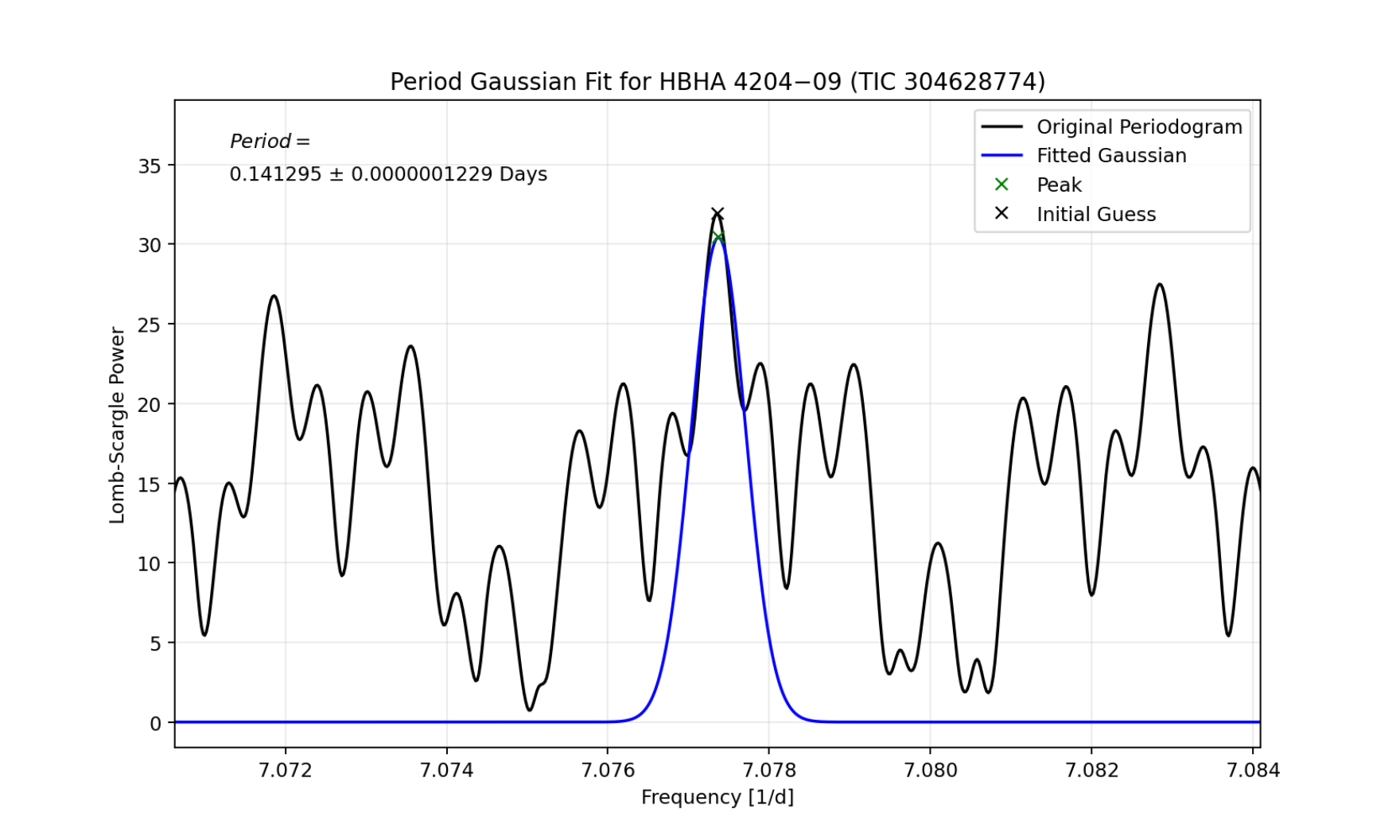}
        \caption{HBHA 4204--09}
    \end{minipage}

    \caption{Periodograms for six of the studied objects. Frequency is in units of inverse days, and amplitude is expressed in Lomb-Scargle Power.}
    \label{fig:periodograms}

\end{figure*}

To find the orbital period of the binary star we applied a Lomb-Scargle periodogram to the residual light curve. Once a periodogram is obtained, we further processed it by applying cubic spline interpolation to smooth the power spectrum, making it easier to identify the desired peaks. Then, we used the smoothed periodogram to detect significant peaks, defined as those exceeding $10\%$ of the maximum power. Once we manually identified the highest peak corresponding to the best-fit period, an initial Gaussian is fitted to the aforementioned peak, based on the peak’s height, center, and the estimated width.  This initial guess for the Gaussian fit is used to find a best-fit Gaussian function allowing for a more accurate peak characterization, including the time of eclipse.  We present an example periodogram fit in Figure \ref{fig:periodogram_exdra} for Ex Dra; the fitted periodograms of the other six sources can be found in Figure \ref{fig:periodograms}.

\subsection{Eclipse time fitting}
\begin{figure}[H]
    \centering
    \includegraphics[width=0.75\linewidth]{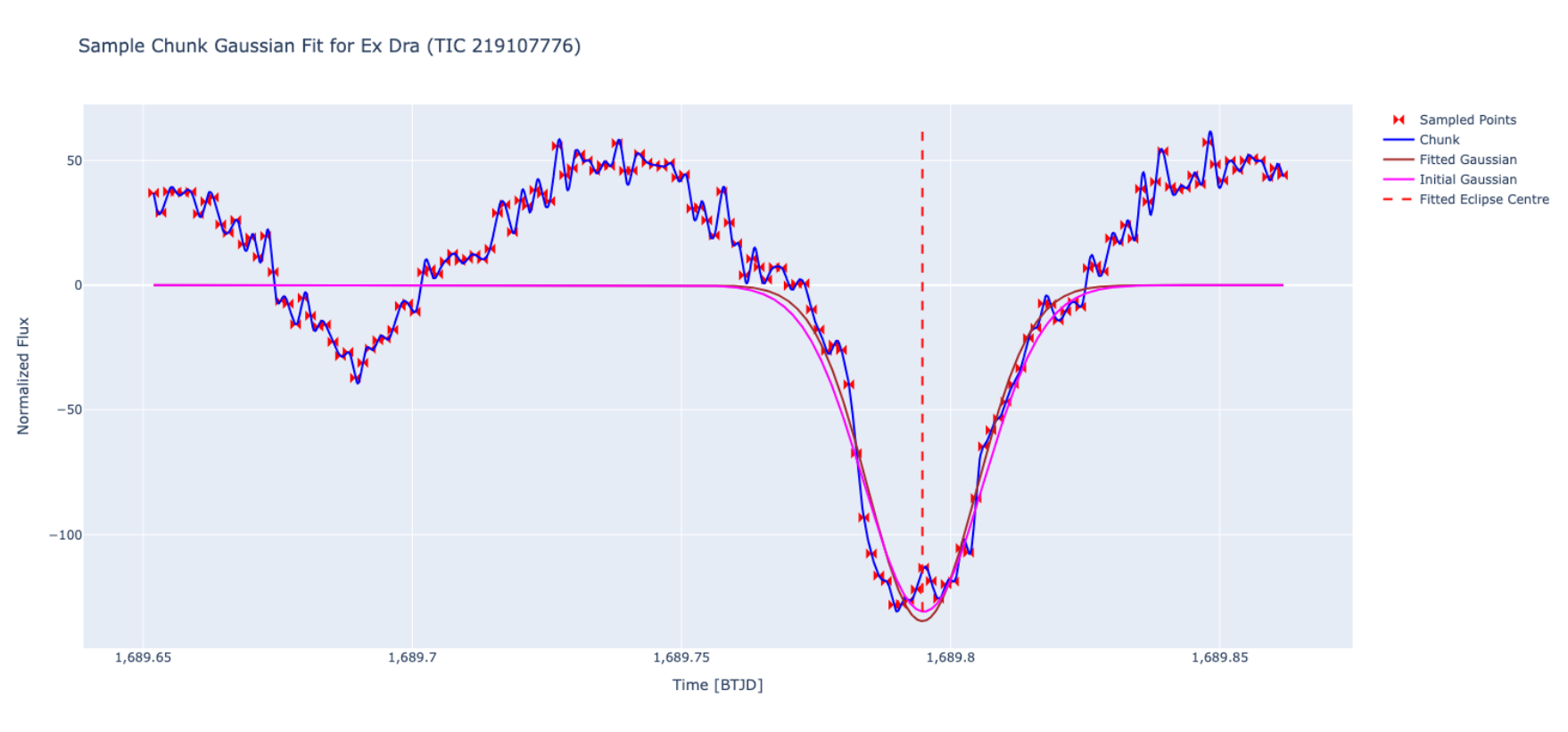}
    \caption{Sample chunk from EX Dra with inverse Gaussian fit. Time is measured in Barycentric Julian Day (BTJD), and amplitude is expressed in normalized flux.}
    \label{fig:gaussian_exdra}
\end{figure}

\begin{figure*}[h!]
    \centering

    \begin{minipage}{0.45\textwidth}
        \centering
        \includegraphics[width=\linewidth]{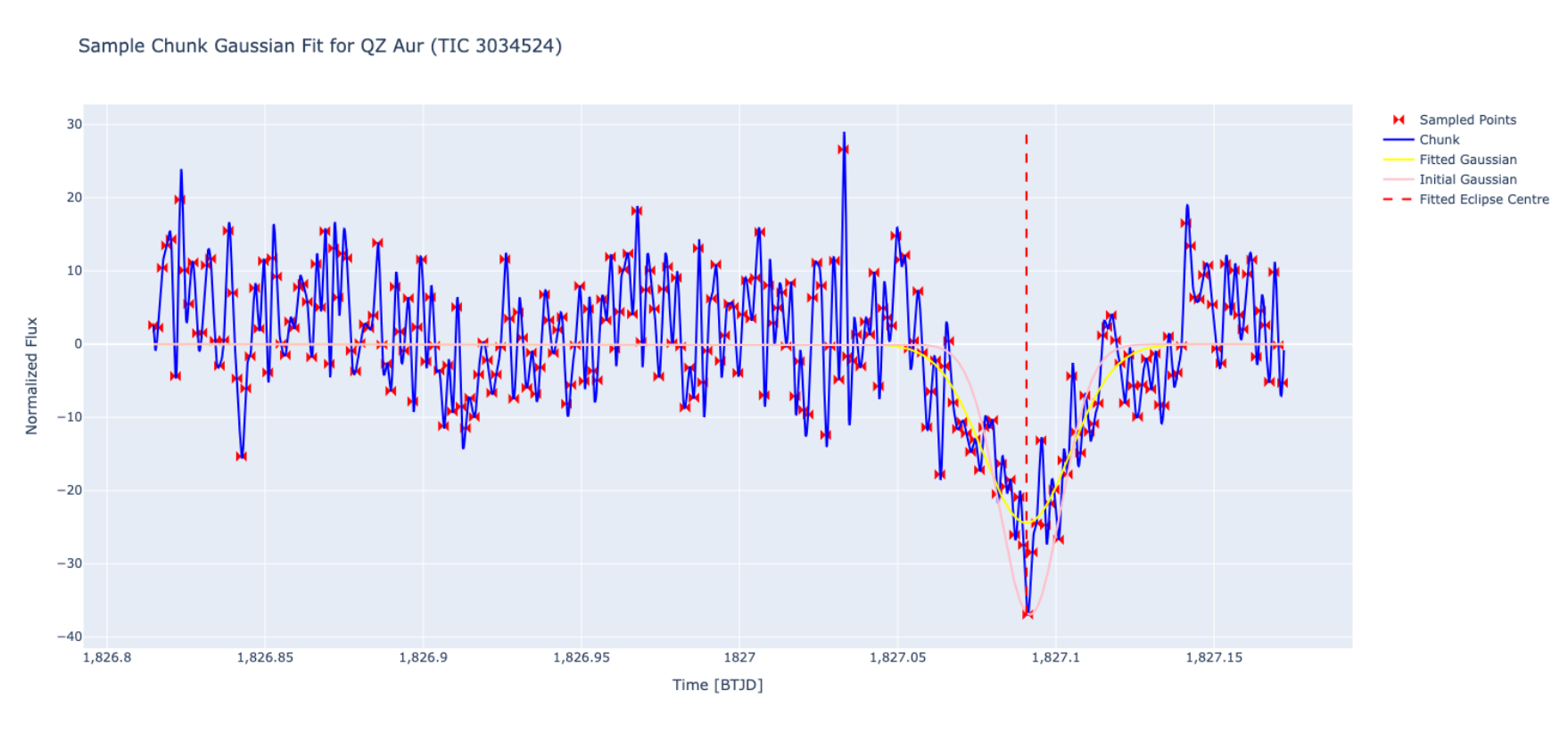}
        \caption{QZ Aur}
    \end{minipage}%
    \hfill
    \begin{minipage}{0.45\textwidth}
        \centering
        \includegraphics[width=\linewidth]{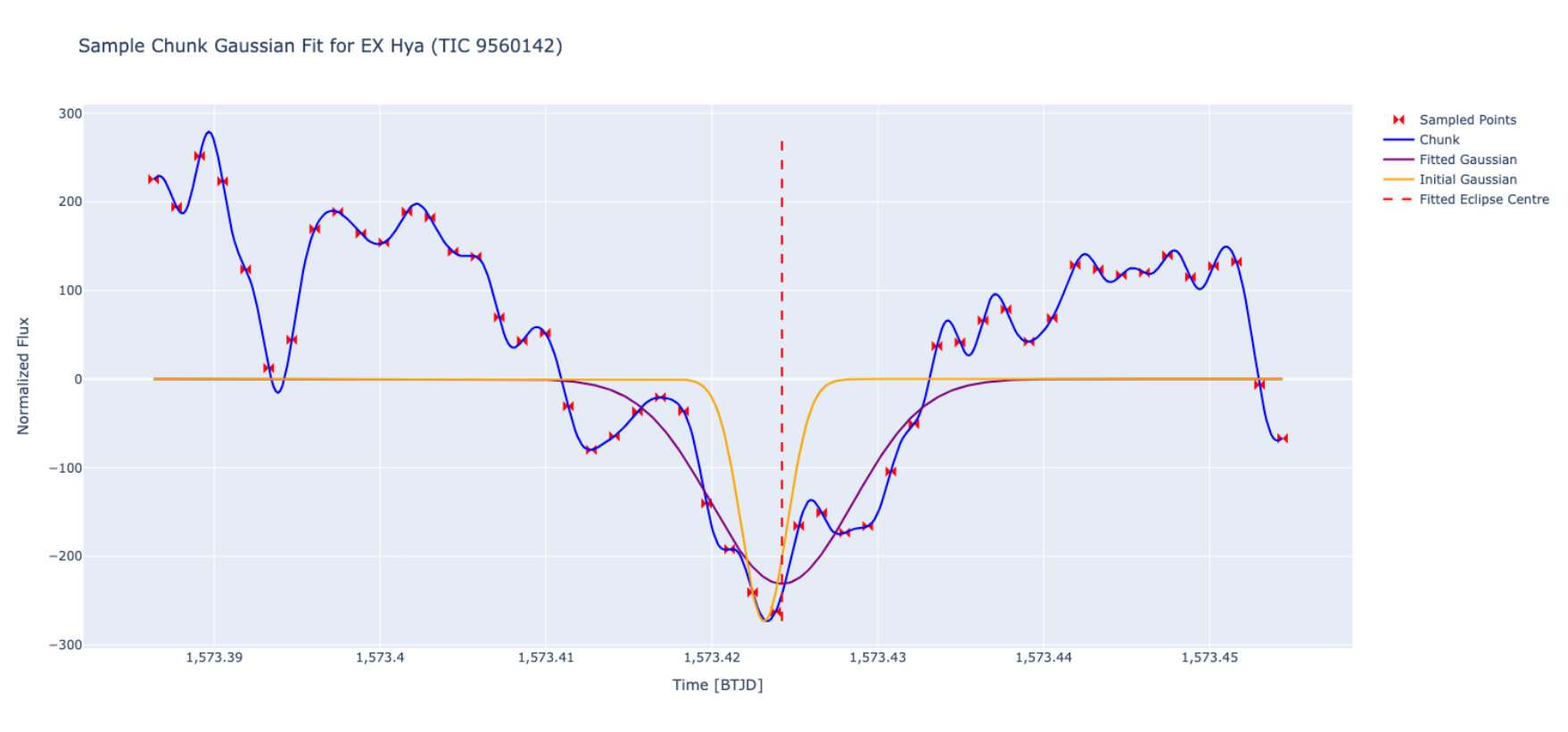}
        \caption{EX Hya}
    \end{minipage}

    \vspace{1em} 

    \begin{minipage}{0.45\textwidth}
        \centering
        \includegraphics[width=\linewidth]{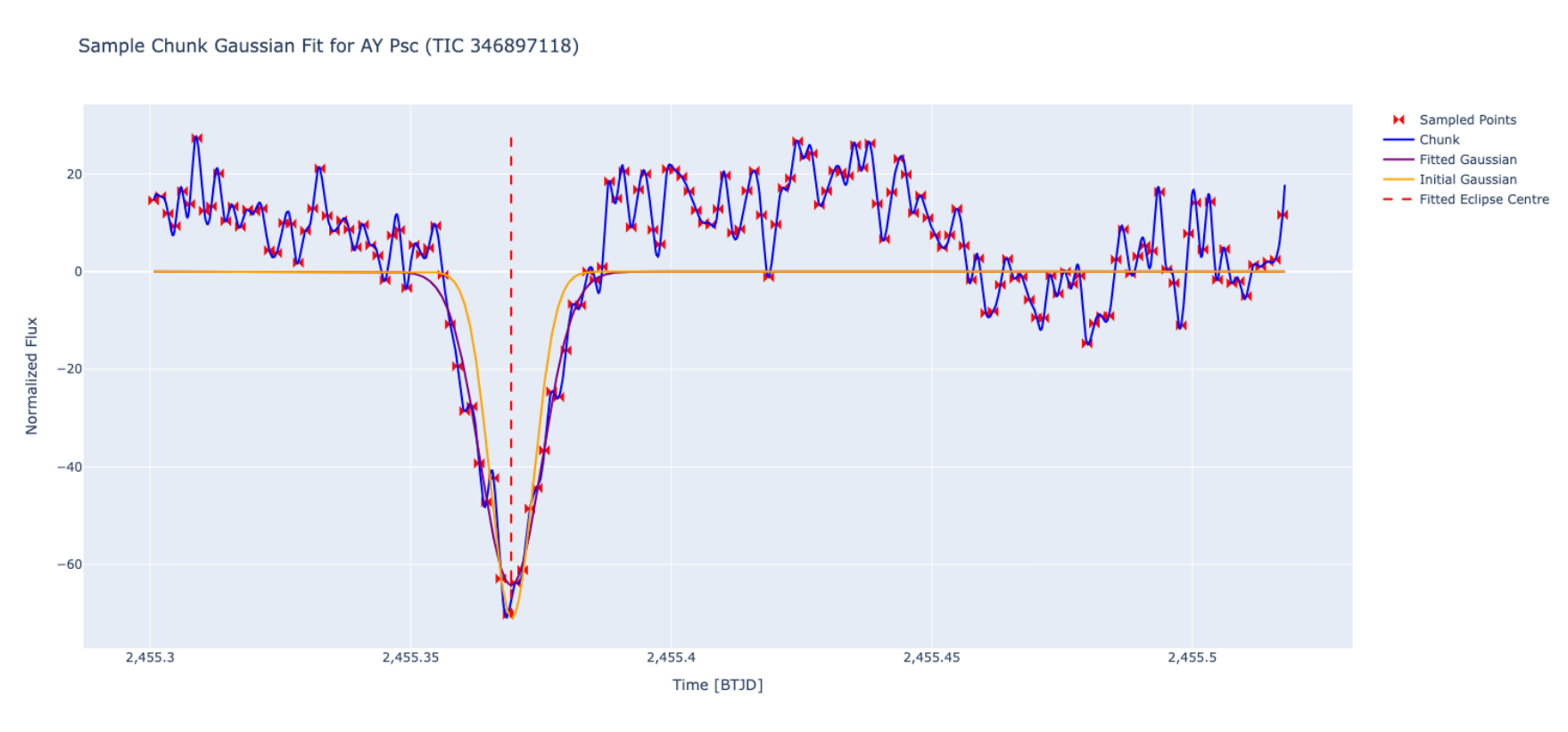}
        \caption{AY Psc}
    \end{minipage}%
    \hfill
    \begin{minipage}{0.45\textwidth}
        \centering
        \includegraphics[width=\linewidth]{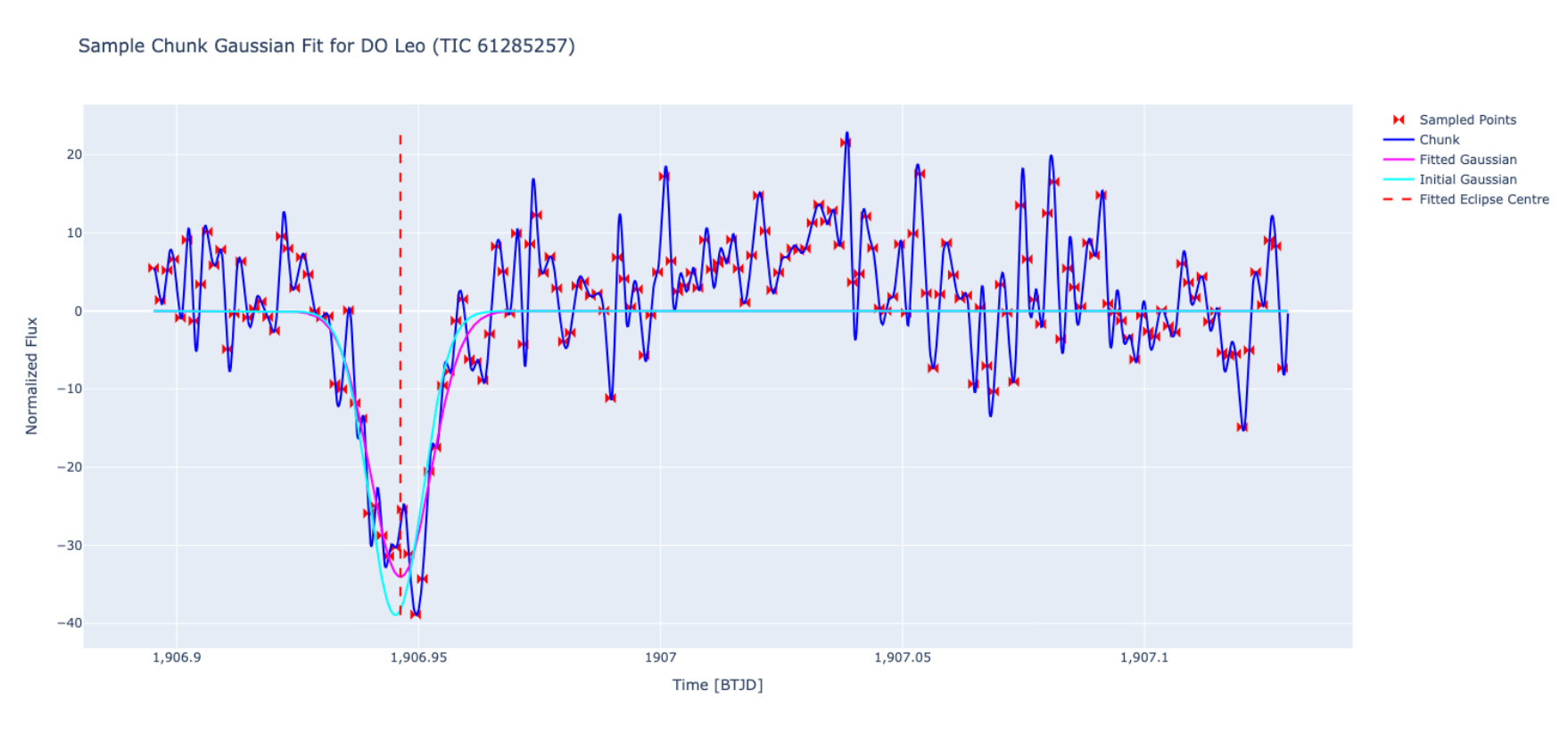}
        \caption{DO Leo}
    \end{minipage}

    \vspace{1em} 

    \begin{minipage}{0.45\textwidth}
        \centering
        \includegraphics[width=\linewidth]{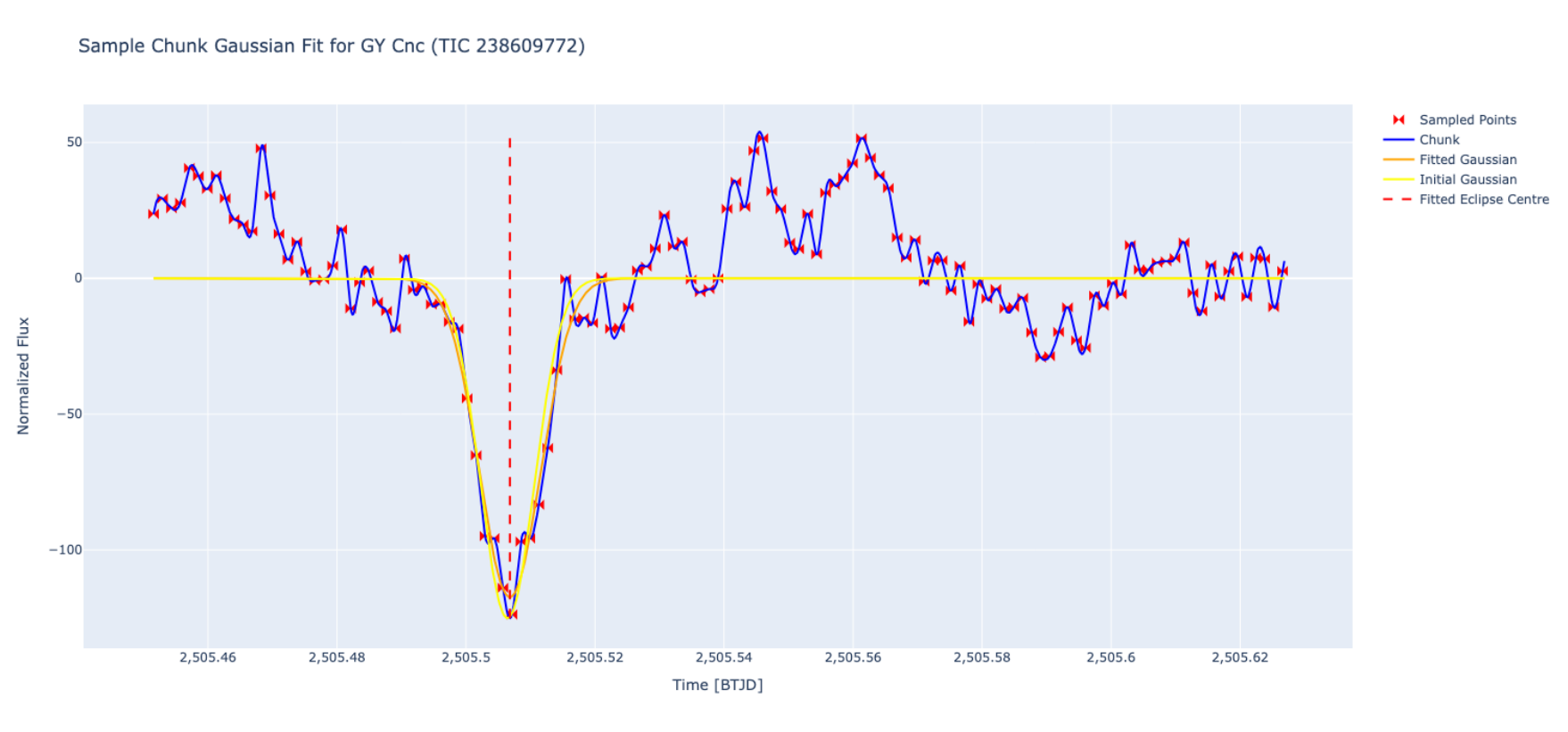}
        \caption{GY Cnc}
    \end{minipage}%
    \hfill
    \begin{minipage}{0.45\textwidth}
        \centering
        \includegraphics[width=\linewidth]{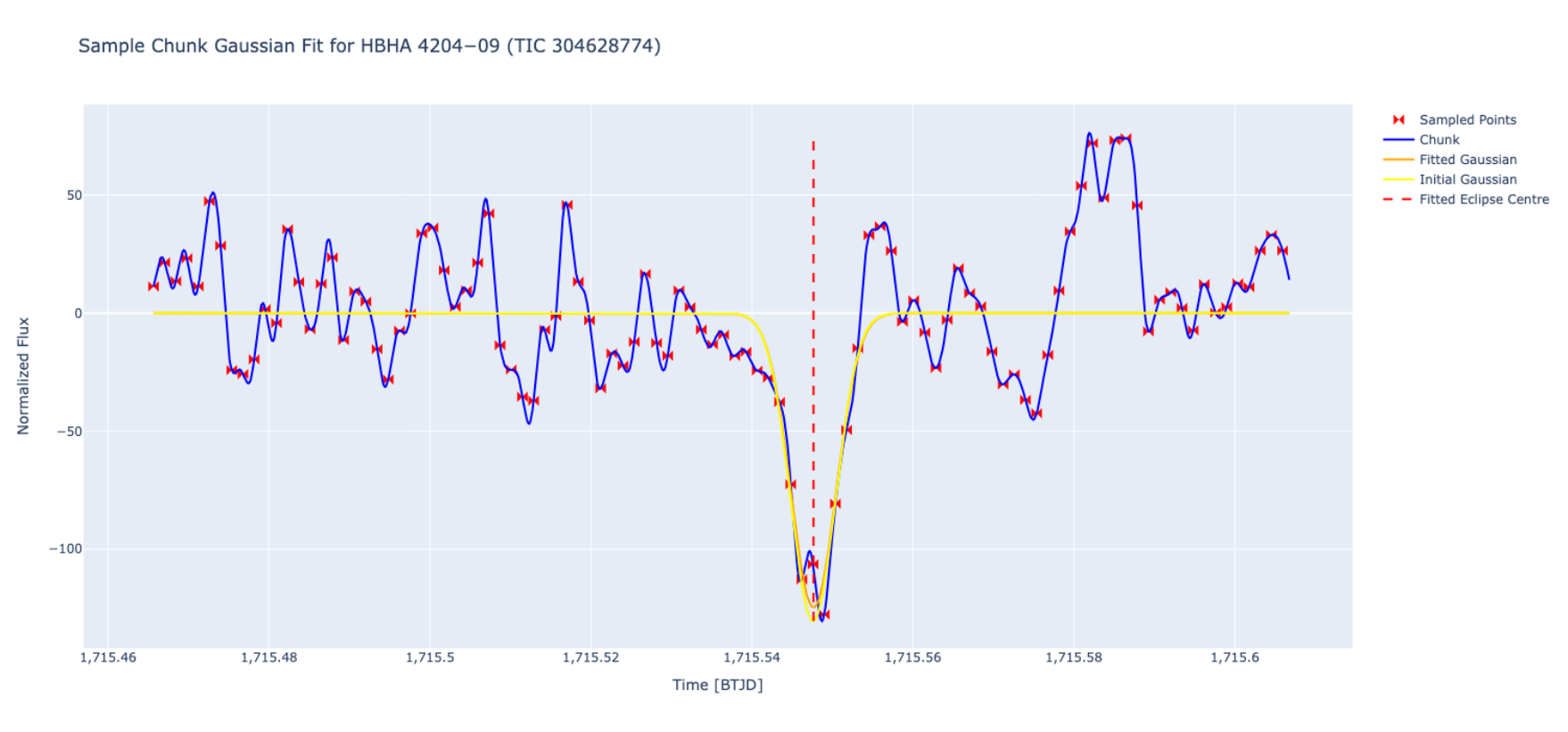}
        \caption{HBHA 4204--09}
    \end{minipage}

    \caption{Sample chunk with eclipse shape and fitted inverse Gaussian for six of the studied objects. Time is measured in Barycentric Julian Day (BTJD), and amplitude is expressed in normalized flux.}
    \label{fig:sample_chunk}

\end{figure*}

Once we determined the orbital period, we used it to divide the light curve data into one period wide chunks containing the primary and secondary eclipses. We modeled the primary eclipse with an inverted Gaussian. From a sample chunk, we manually select two points in the eclipse portion of the light curve for which an initial guess of the full width at half maximum (FWHM) and the center of the eclipse are derived. Once the initial guess for both parameters are determined, we use them to generate a best-fit Gaussian for all eclipses in the light curve. An inverted Gaussian appears to work well for sources like EX Dra (Figure \ref{fig:gaussian_exdra}); however, we note here that low signal-to-noise, and contamination from features associated with outbursts and the accretion disk, can make fitting difficult ($e.g.$, EX Hya in Fig. \ref{fig:sample_chunk}($b$)). As we progress toward surveying additional sources in the future, we may utilize a phase-folded light curve as a template for fitting eclipse times.  This processing step allows us to extract the observed ($O$) eclipse times necessary for creating an $O-C$ diagram. 

\subsection{O-C Diagrams and Period Derivation}
\begin{figure}[H]
    \centering
    \includegraphics[width=0.75\linewidth]{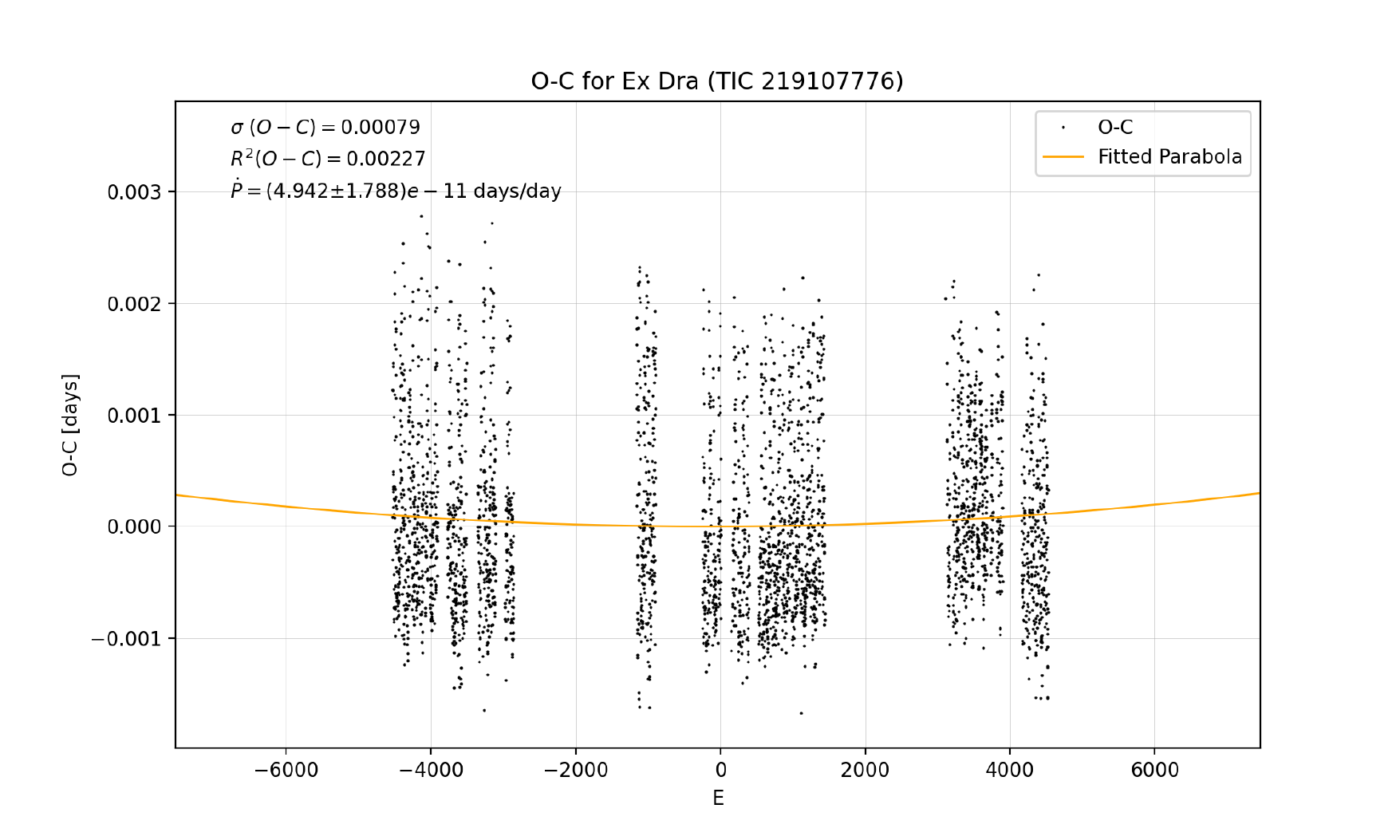}
    \caption{$O-C$ for EX Dra with time measured in epochs.}
    \label{fig:O_C_exdra1}
\end{figure}

After extracting the observed eclipse times ($O$), we then calculated the predicted ($C$) eclipse times from the following equation:
$$C=T_0+EP$$ 
where $T_0$ is the initial time value in the light curve, $E$ is the epoch or cycle number of the eclipse, and $P$ is the orbital period we determined from the Lomb-Scargle periodogram. Once we have both $O$ and $C$ values, we plot $O-C$ versus eclipse cycle number $(E)$ and fit a quadratic curve to the data. If there is a non-zero period derivative $\dot{P}$, we expect the $O-C$ diagram to exhibit a quadratic curvature according to the following relationship:
$$O-C=\frac{1}{2}P\dot{P}E^2 .$$
The $O-C$ diagram for EX Dra (Fig. \ref{fig:O_C_exdra1}) exhibits a positive quadratic curve, indicating a positive value of $\dot{P}$.  The $O-C$ diagrams of the other six sources exhibit positive and negative quadratic trends (see Fig. \ref{fig:o-c}).

\section{Results and Discussion}

We summarize the results of our time series analysis of the target sources in Table \ref{tab:cv_p_pdot} and briefly discuss our findings for each source. The periods are listed in units of days, and the period derivative $\dot{P}$ is in units of days/day.  

\begin{table}[H]
    \centering
    \setlength{\tabcolsep}{15pt}  
    \begin{tabular}{ccc}
        \hline
        \hline
        \textbf{CV} & $\mathbf{P}$ \textbf{(days)} & $\mathbf{\dot{P}} \, (10^{-9} \text{ days/day})$ \\
        \hline
        \hline
        QZ Aur       & 0.35750347$\pm$0.0000001 & 1.563$\pm$0.198 \\
        EX Dra       & 0.209937 $\pm$ 0.00000000795    & 4.942$\pm$1.788$\times 10^{-2}$ \\
        EX Hya       & 0.068228 $\pm$ 0.000000021   & 9.1322$\pm$4.142$\times 10^{-1}$ \\
        AY Psc       & 0.217323$\pm$0.0000008643     & 2.189 $\pm$1.388 \\
        DO Leo       & 0.234516 $\pm$ 0.0000000555     & -5.901$\pm$0.875$\times 10^{-1}$ \\
        GY Cnc       & 0.175441$\pm$0.0000005915     & -1.912$\pm$3.930$\times 10^{-1}$ \\
        HBHA 4204--09 & 0.141295$\pm$0.0000001229   & -7.279$\pm$5.059$\times 10^{-2}$ \\
        \hline
        \hline
    \end{tabular}
    \caption{$P$ and $\dot{P}$ values for our sources.}
    \label{tab:cv_p_pdot}
\end{table}
\subsection{QZ Aur}
QZ Aur is classified as a classical nova, last seen to go nova in 1964 \citep{sanduleak}. In its quiescent state, QZ Aur has a typical $V$ magnitude of $~17$ \citep{sheng-Bang}. The orbital period of QZ Aur is $0.3575$ days, or about $8.58$ hours, in agreement with previous measurement \citep{allen}.  Using only TESS data, our $O-C$ diagram (Fig. \ref{fig:o-c}$(a)$) yielded a positive $\dot{P}$ of $1.563 (\pm 0.198)\times 10^-9$.  However, after including additional eclipse times over a span of several decades, \cite{schaefer2024} found that the $O-C$ is inverted, with a $\dot{P}$ of $-3.9(\pm 1.4)\times 10^{-11}$. While the cadence and repetition of TESS data can provide us with highly precise orbital information on short time scales of a few years, we cannot discount the power of time domain astronomy to reflect the true nature of the orbital dynamics of complex systems like cataclysmic variables.

\subsection{EX Dra}
EX Dra was discovered as a variable star in the Hamburg Quasar Survey \citep{bade1989}. It was identified as an eclipsing dwarf nova due to its periodic outbursts and the accumulation and release of material from the donor star onto the white dwarf \citep{barwig1993}. During outbursts, EX Dra’s brightness can rise from its quiescent state of magnitude of 15 to as high as 13.5. Outbursts occur every 10--30 days and last about 10 days \citep{voloshina2021}. The orbital period of EX Dra is 0.21 days, or about 5 hours, in agreement with previous observations \citep{voloshina2021}. Using only TESS data, a fit of our $O-C$ diagram (Fig. \ref{fig:O_C_exdra1}) shows a $\dot{P}$ of $4.942 (\pm 1.788) \times 10^{-11})$, almost one standard deviation from the value determined by \cite{schaefer2024} of $3.05 (\pm 0.32)\times 10^{-11}$, who used data from TESS and eclipse epochs from as early as 1991.

\subsection{EX Hya}
EX Hya was discovered as a variable star by \cite{brun} and classified as a dwarf nova in 1980 \citep{Andronov_2013}. It is an eclipsing intermediate polar-type cataclysmic variable with ultrashort periods.  EX Hya shows dwarf nova oscillations with highly erratic outbursts, where the magnitude can vary from about $14.1$ to $9.2$ \citep{AAVSO2024}. This eclipsing binary has a very short orbital period of $0.068234$ days, or about $1.6$ hours, in agreement with the optical and X-ray observations of \cite{Mukai2024}. The $\dot{P}$ derived from our $O-C$ diagram (Fig. \ref{fig:o-c}$(b)$) is $9.132 (\pm 4.142) \times 10^{-10})$. However, over a longer time period, \cite{schaefer2024} found a $\dot{P}$ of $- 7.2 (\pm 0.5)\times 10^{-12}$ from records dating back to 1962. This CV is certainly more of a challenge to characterize, as the $O-C$ timing jitter require longer term time domain observations to characterize its orbital dynamics.  

\subsection{AY Psc}
AY Piscium is classified as a Z Camelopardalis (Z Cam) type dwarf nova, identified by \cite{green82}. This subtype of cataclysmic variable stars exhibits both normal outbursts and standstills, where the brightness remains nearly constant for extended periods \citep{K_ra_2023}. AY Psc experiences quasi-periodic outbursts with an amplitude of about 2.5 magnitudes during which the system reaches the maximal brightness of 14.6 mag \citep{Han_2017}. For AY Psc, we measured an orbital period of 0.2173 days, or about 5.2 hours, in agreement with the period found by \cite{K_ra_2023}. The $\dot{P}$ we measured from a quadratic fit of the $O-C$ diagram (Fig. \ref{fig:o-c}$(c)$) is $2.189 (\pm 1.388)\times 10^{-9}$; however we note here that \cite{K_ra_2023} found a value about 100 times lower ($2.08\times 10^{-11}$) from recorded eclipse times spanning almost thirty years.

\subsection{DO Leo}
DO Leo was found by \cite{green86} while searching for faint blue stars at high galactic latitudes. This binary system is a dwarf nova of the SU Ursae Majoris (SU UMa) subtype with a magnitude of 16 \citep{Abbott_1990}. DO Leo undergoes frequent outbursts, with super outbursts occurring irregularly during its orbital period of 0.2345 days, or about 5.63 hours, as found by \cite{Abbott_1990} and \cite{bruch24}. The $\dot{P}$ derived from our $O-C$ diagram (Fig. \ref{fig:o-c}$(d)$) is $-5.90 (\pm 0.88)\times 10^{-10}$.

\subsection{GY Cnc}
GY Cnc is an eclipsing cataclysmic variable star with an M-dwarf donor of spectral type M3 \citep{warwick73257}. The system experiences frequent dwarf nova outbursts and eclipse timings with significant variability in brightness between $11.4$ and $17.94$ magnitudes \citep{BAA2024}. The orbital period we obtained for this object is 0.1754, or about 4.2 hours, in agreement with studies by \cite{Littlefair2023} and \cite{canizares18}. The $\dot{P}$ derived from our $O-C$ diagram (Fig. \ref{fig:o-c}$(e)$) is $-1.912(\pm 3.930) \times 10^{-10}$. Using Kepler observations, \cite{canizares18} were unable to measure a $\dot{P}$.  Our value has a high error, likely because there were only two TESS segments used in the quadratic fit. Future TESS (or other) observations can more precisely constrain the period derivative for this system.

\subsection{HBHA 4204 -- 09}
\begin{figure*}[h!]
    \centering

    \begin{minipage}{0.45\textwidth}
        \centering
        \includegraphics[width=\linewidth]{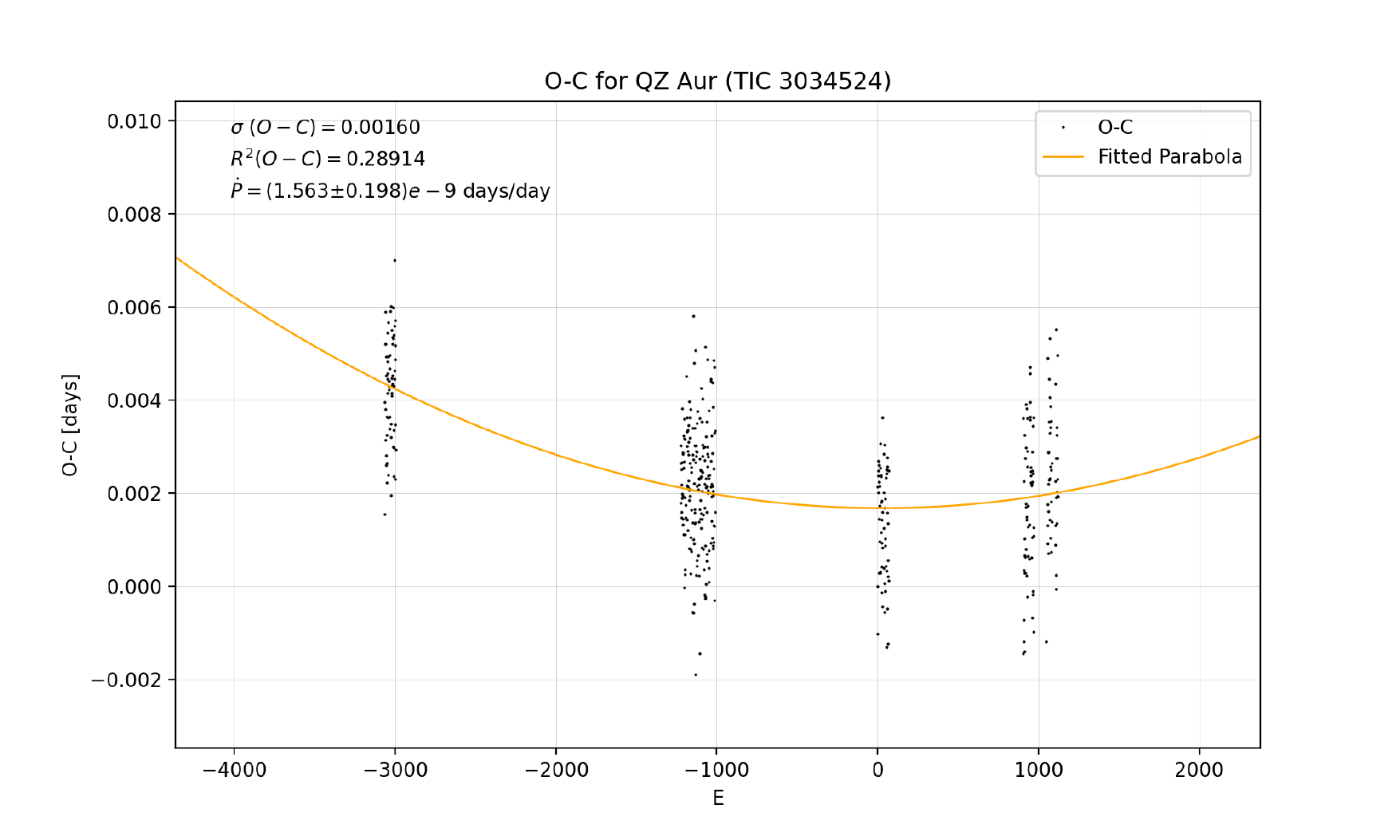}
        \caption{QZ Aur}
    \end{minipage}%
    \hfill
    \begin{minipage}{0.45\textwidth}
        \centering
        \includegraphics[width=\linewidth]{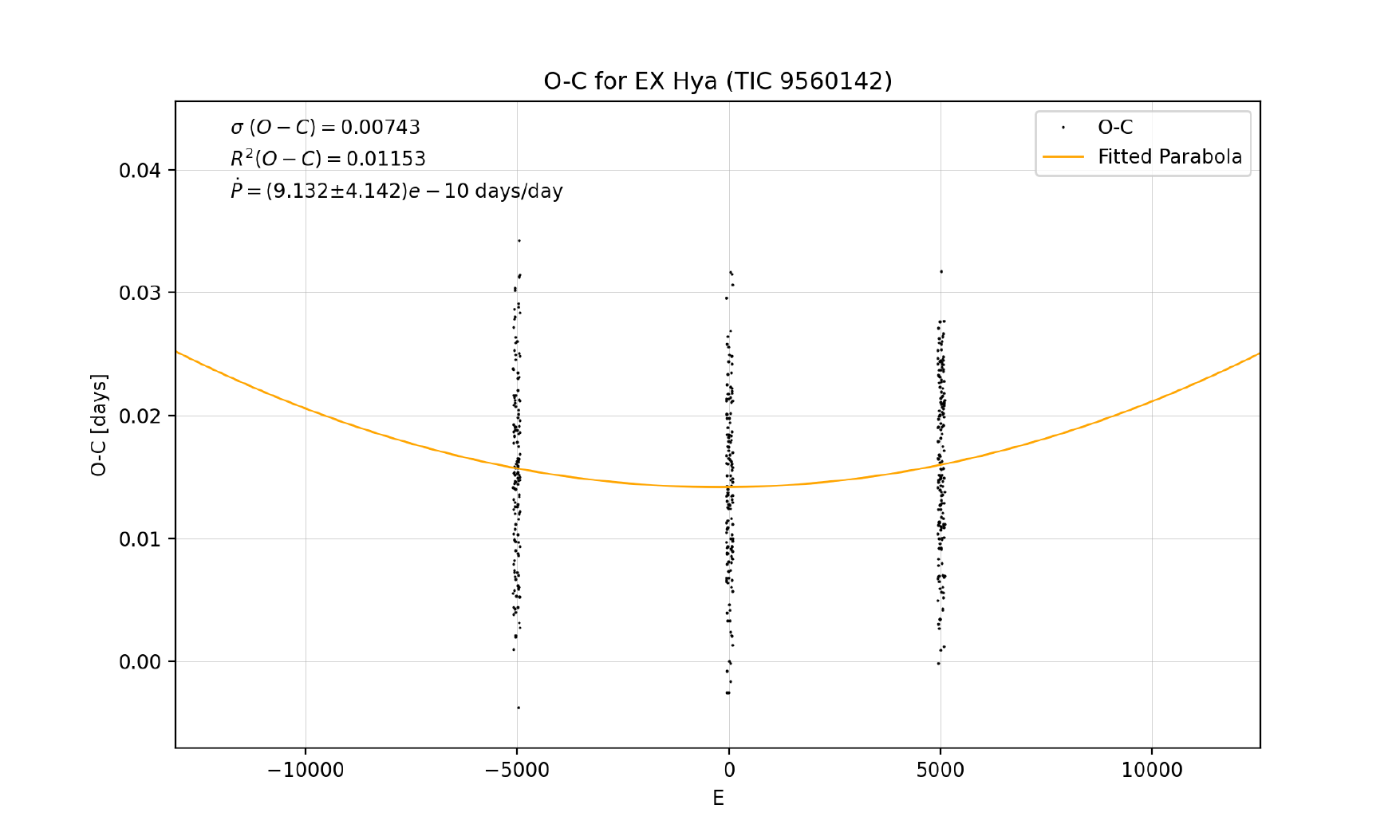}
        \caption{EX Hya}
    \end{minipage}

    \vspace{1em} 

    \begin{minipage}{0.45\textwidth}
        \centering
        \includegraphics[width=\linewidth]{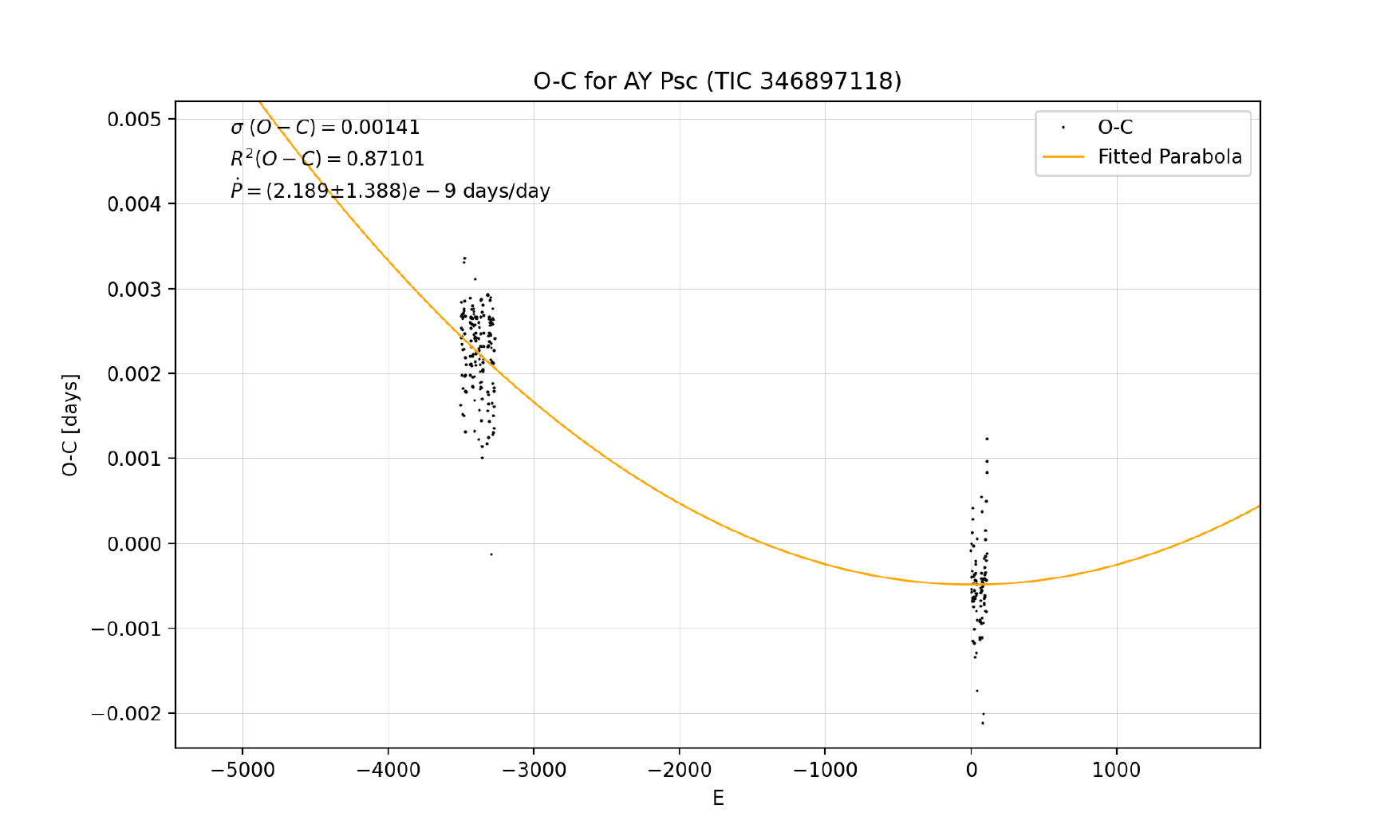}
        \caption{AY Psc}
    \end{minipage}%
    \hfill
    \begin{minipage}{0.45\textwidth}
        \centering
        \includegraphics[width=\linewidth]{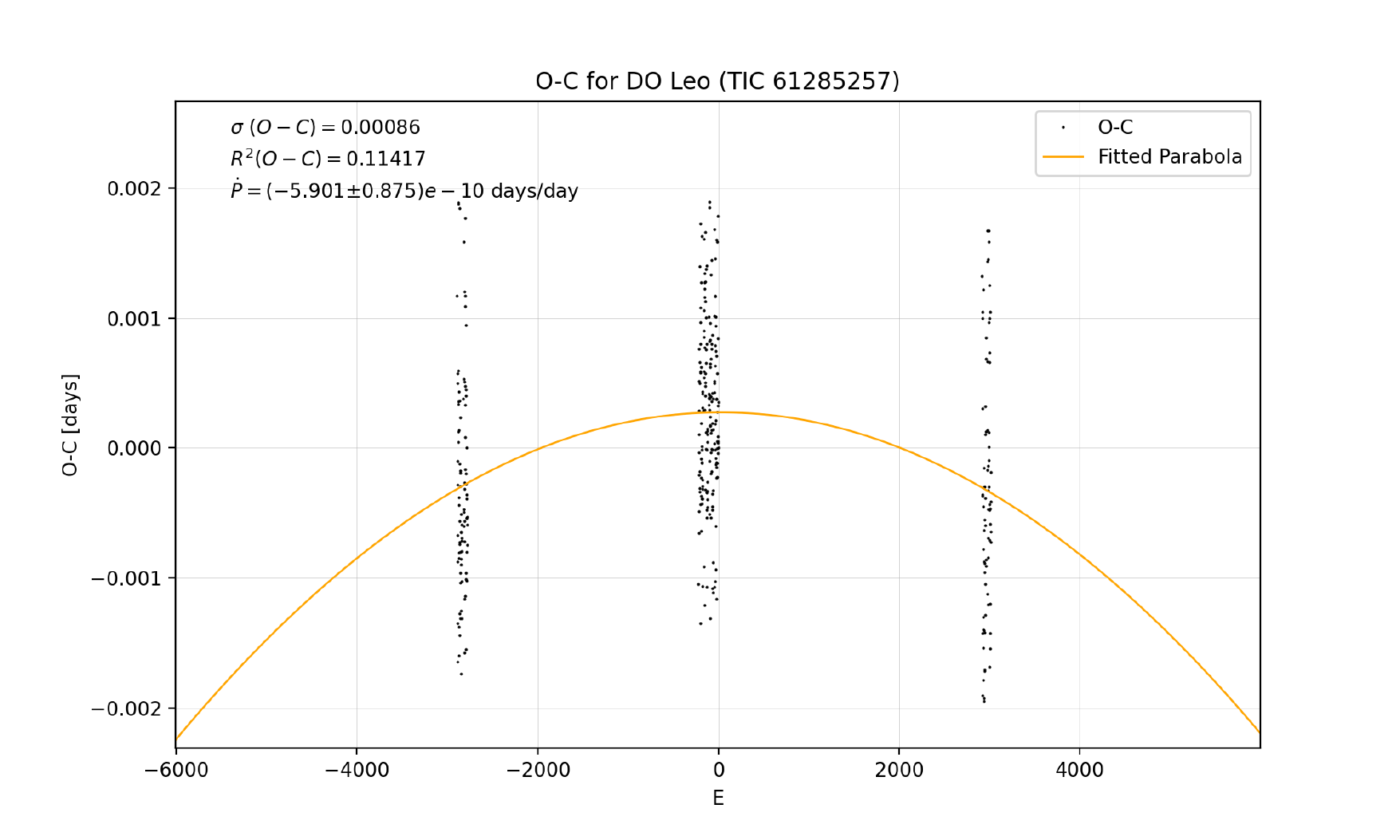}
        \caption{DO Leo}
    \end{minipage}

    \vspace{1em} 

    \begin{minipage}{0.45\textwidth}
        \centering
        \includegraphics[width=\linewidth]{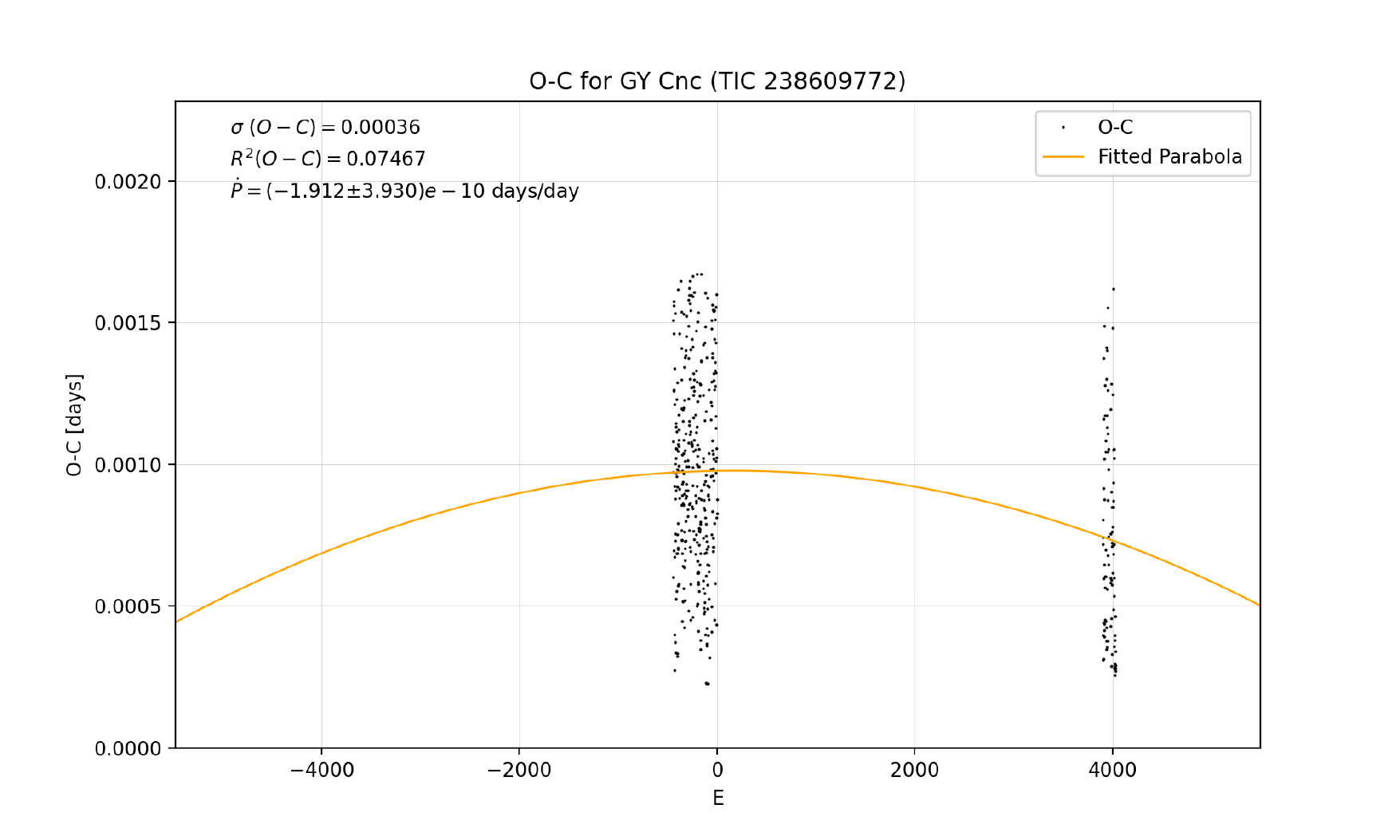}
        \caption{GY Cnc}
    \end{minipage}%
    \hfill
    \begin{minipage}{0.45\textwidth}
        \centering
        \includegraphics[width=\linewidth]{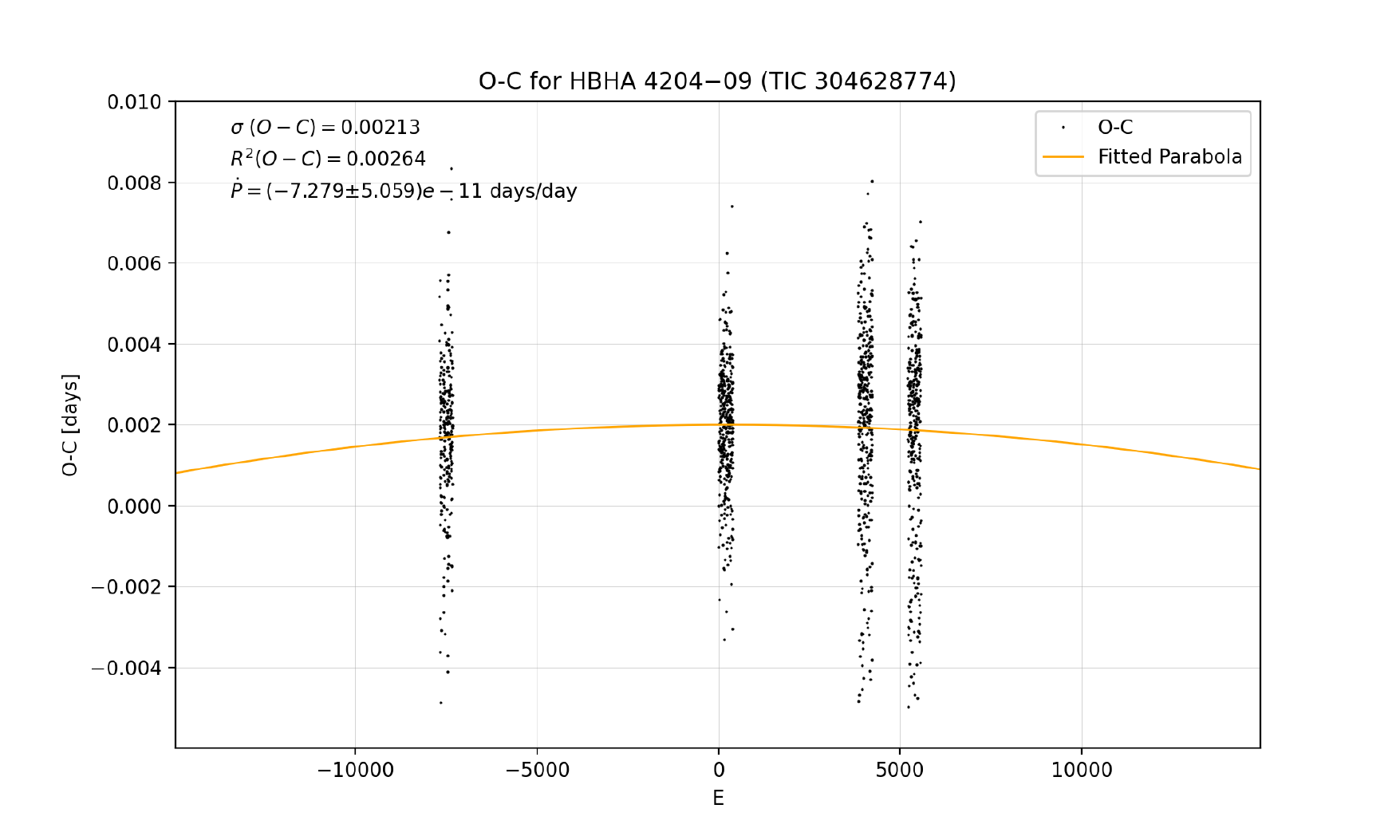}
        \caption{HBHA 4204--09}
    \end{minipage}

    \caption{O-C diagrams with time measured in epochs for the six studied objects.}
    \label{fig:o-c}

\end{figure*}

HBHA 4204 -- 09 is an eclipsing binary system with negative superhumps (nSHs) detected in its light curve. It was discovered by ASAS-SN and classified as a CV by \cite{jayasinghe} with a mean magnitude of 16.9. This system has been studied more extensively in recent years, particularly with data from the TESS. The orbital period of HBHA 4204--09 is approximately 0.1413 days, or about 3.4 hours, which is the same found by \cite{Stefanov_2023}. The $\dot{P}$ derived from our $O-C$ diagram (Fig. \ref{fig:o-c}$(f)$) is $-7.279(\pm 5.059)\times 10^{-11}$. While the error is high compared to our $\dot{P}$ value, it is almost a factor of 10 lower than the error for GY Cnc because there are more segments to constrain the quadratic fit.

\section{Conclusion}

We have analyzed the TESS light curves of seven CV systems by creating Python code to download two-minute cadence photometric flux data, process them to find the orbital periods of these binary systems, and compared the observed to predicted ($O-C$) eclipse times to search for positive or negative changes of period with time. $O-C$ diagrams are a versatile tool in astrophysics, particularly beneficial for studying binary star systems, including eclipsing cataclysmic variable stars, which system we can observe edge-on. Continued studies of these high-inclination systems could aid us in understanding the complex dynamics and evolutionary processes of these systems.\\

The fitted orbital periods of all our eclipsing CVs are in agreement with previous measurements found in literature. However, our success with determining the period derivative varied. By using only TESS data, our $O-C$ diagrams had a three to four year time range from which we could make quadratic fits for $\dot{P}$.  Our measured values of $P$ and $\dot{P}$ for EX Dra were within one sigma of previous measurements, confirming the veracity of the processing applications we developed.  However, our $\dot{P}$ values for QZ Aur, EX Hya, and AY Piscium did not agree with those determined from $O-C$ studies that included data over longer time intervals. Our results can be improved with the inclusion of additional eclipse epochs for our sample CVs along with other systems we chose to study in the near future.  

Our most promising results were those for DO Leo, GY Cnc, and HBHA 4204--09.  Our measured periods are in agreement with those found in literature, and there appears to be no recent determination of $\dot{P}$ in literature for comparison.  All three sources have negative period derivatives, indicating that their orbital periods are decreasing over time. A commonly-cited explanation for period decrease over time is the action of magnetic braking in the binary system. 

\section{Acknowledgments}
The authors acknowledge support from the Furman University Summer Research Fellowship program.
This paper includes data collected by the TESS mission. Funding for the TESS mission is provided by
NASA’s Science Mission Directorate. This research has used the SIMBAD database, operated at
CDS, Strasbourg, France. We acknowledge with thanks the variable star observations from the AAVSO International Database contributed by observers worldwide and used in this research.

\section{Appendix}
\vspace{0.5em}
\subsection{Light Curves} 
 
\begin{figure*}[h!]
    \centering

    \begin{minipage}{0.45\textwidth}
        \centering
        \includegraphics[width=\linewidth]{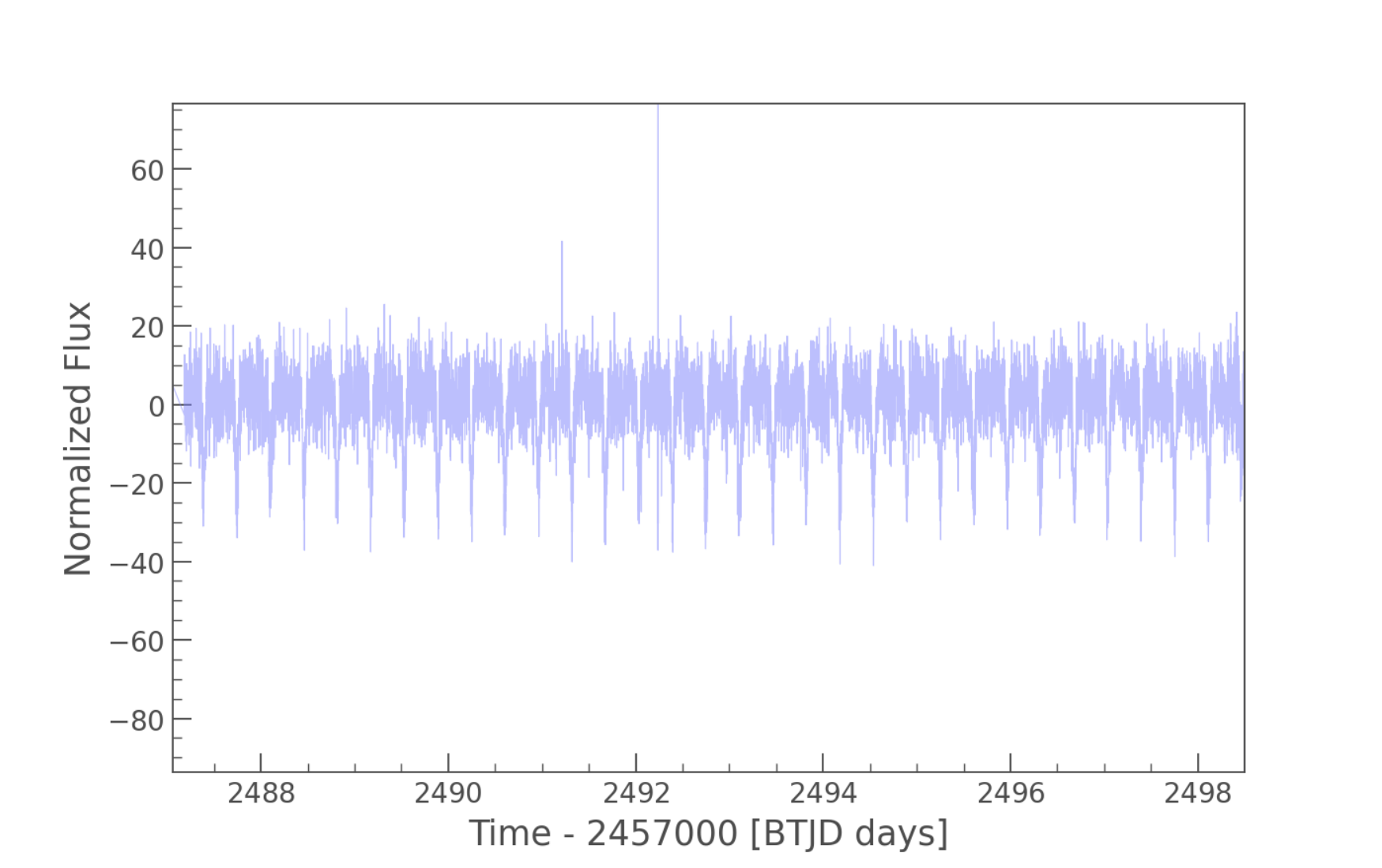}
        \caption{QZ Aur}
    \end{minipage}%
    \hfill
    \begin{minipage}{0.45\textwidth}
        \centering
        \includegraphics[width=\linewidth]{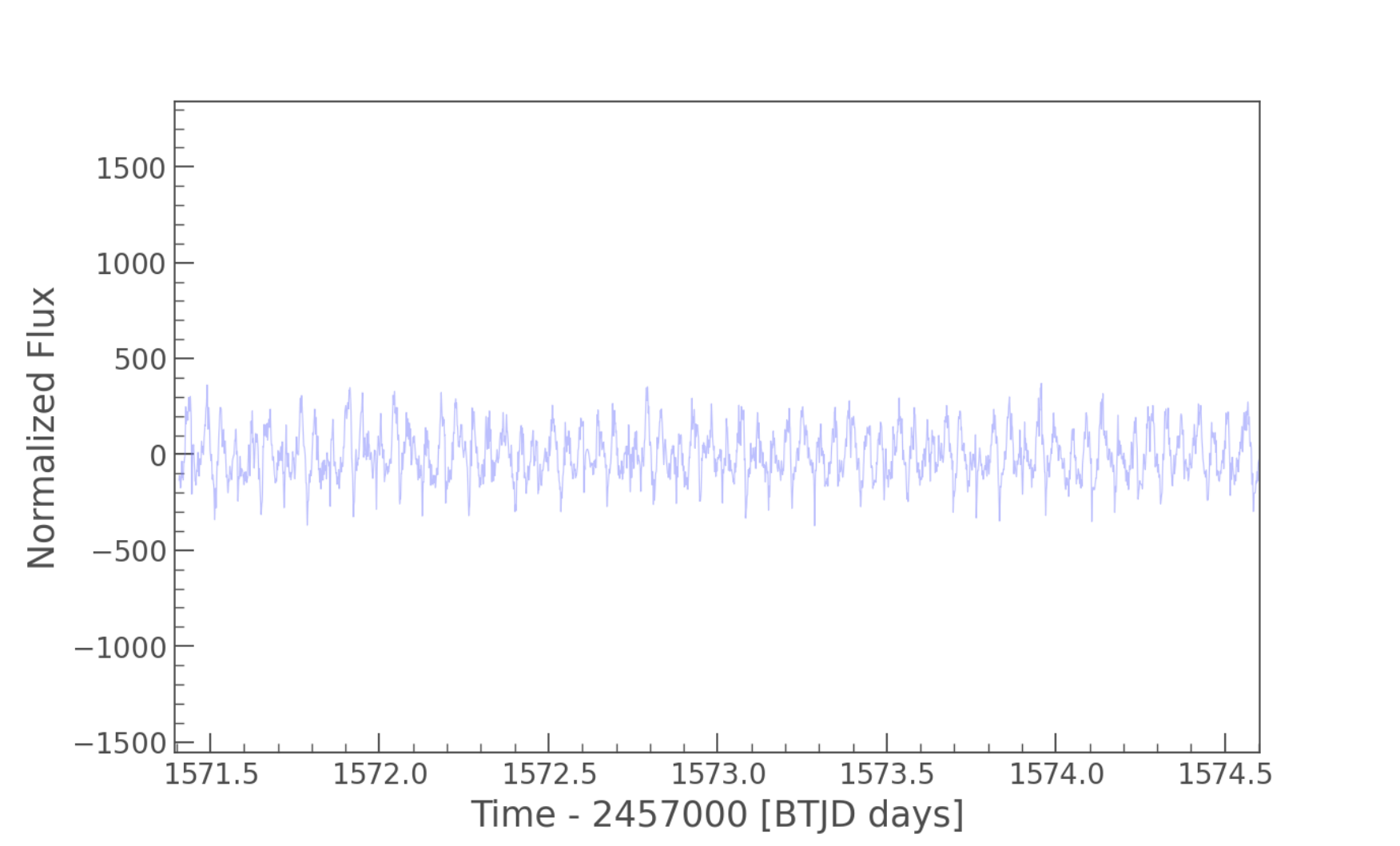}
        \caption{EX Hya}
    \end{minipage}

    \vspace{1em} 

    \begin{minipage}{0.45\textwidth}
        \centering
        \includegraphics[width=\linewidth]{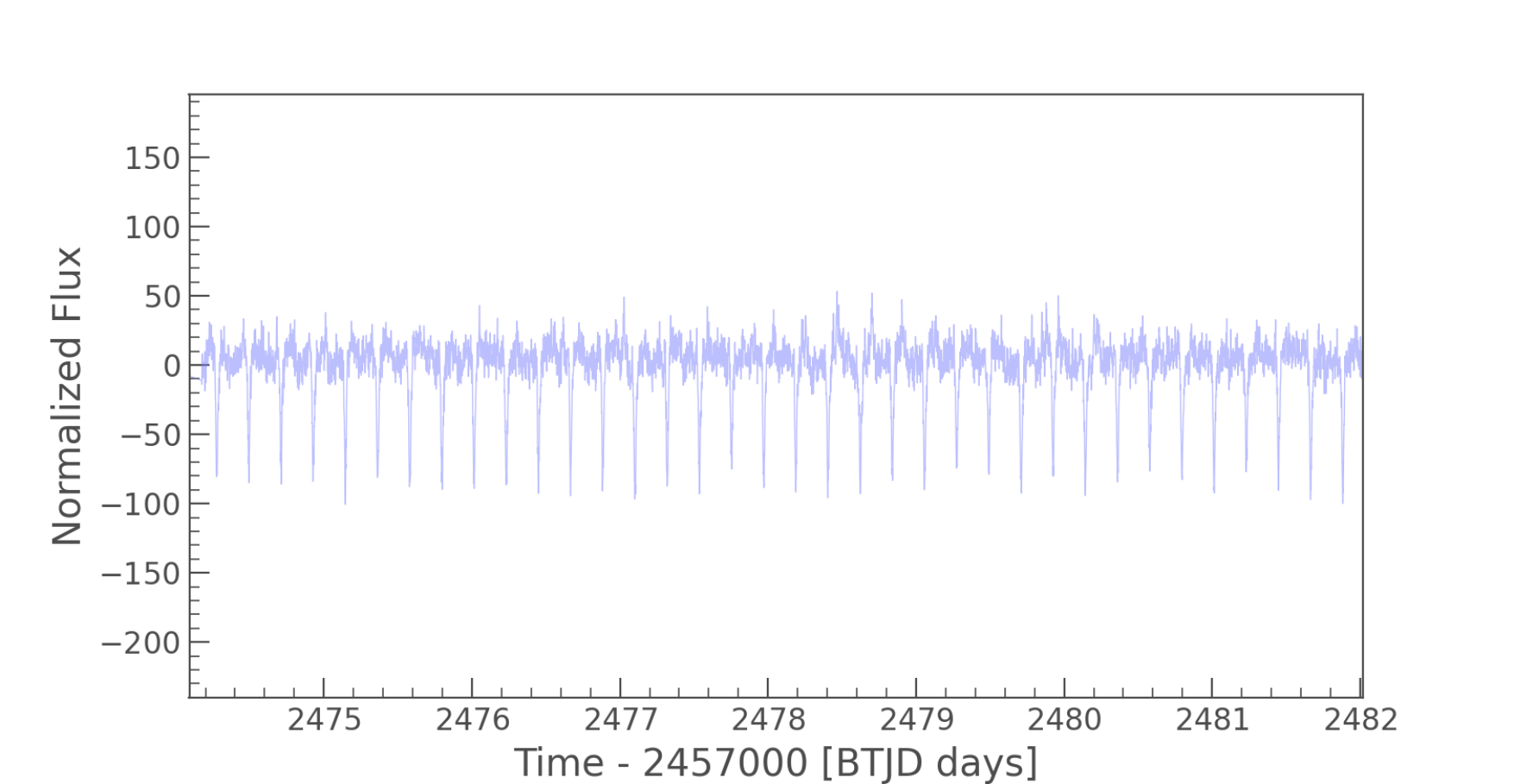}
        \caption{AY Psc}
    \end{minipage}%
    \hfill
    \begin{minipage}{0.45\textwidth}
        \centering
        \includegraphics[width=\linewidth]{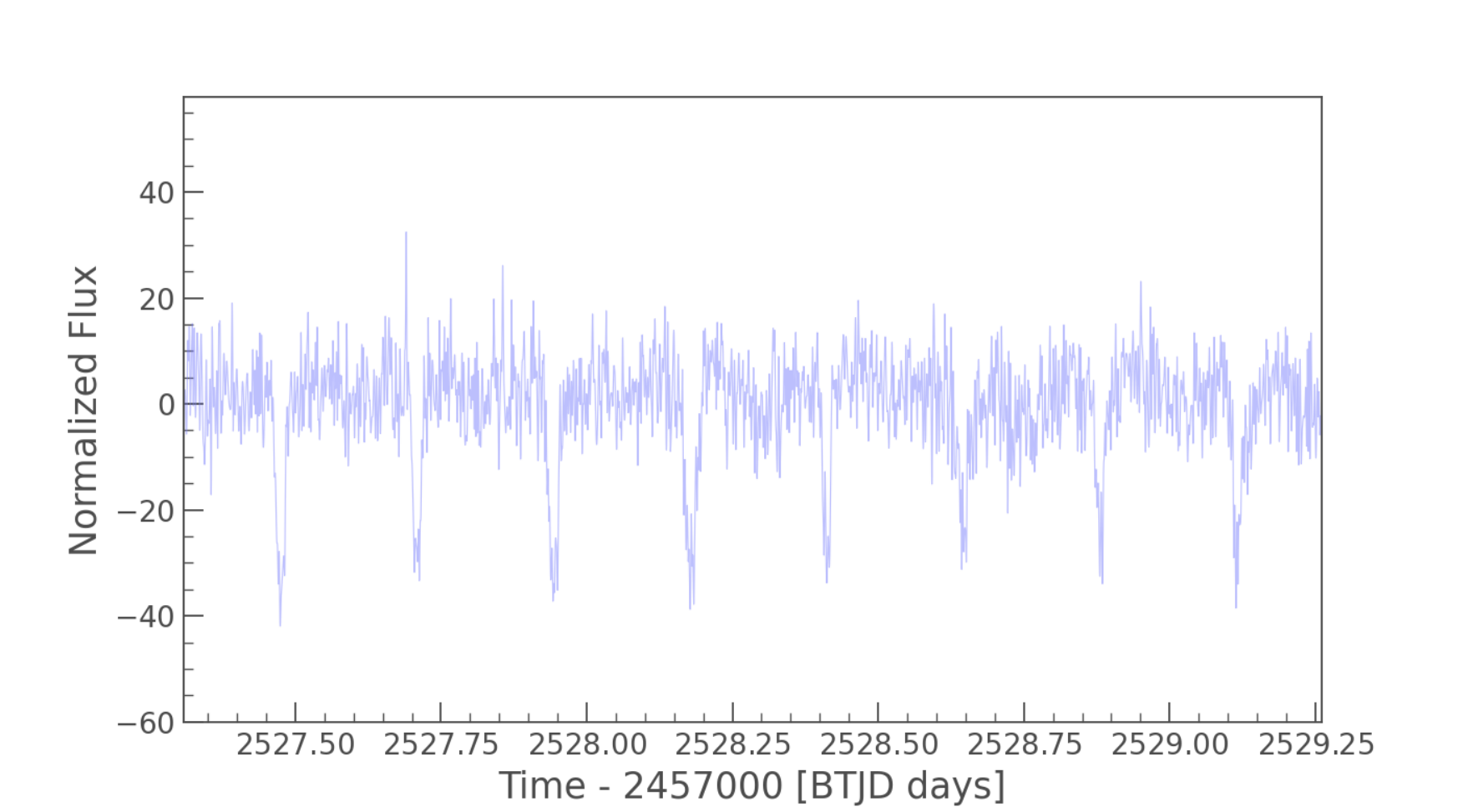}
        \caption{DO Leo}
    \end{minipage}

    \vspace{1em} 

    \begin{minipage}{0.45\textwidth}
        \centering
        \includegraphics[width=\linewidth]{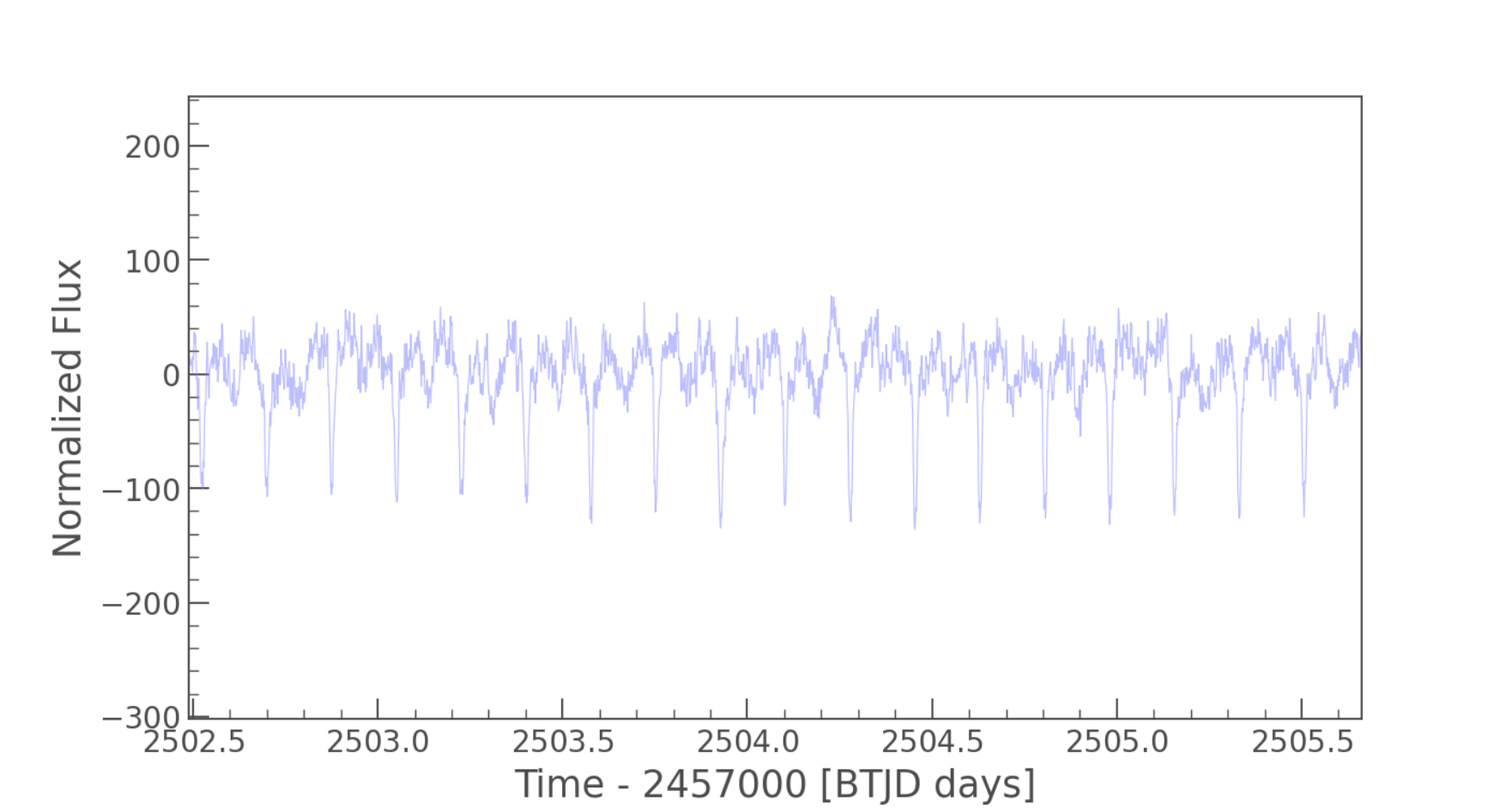}
        \caption{GY Cnc}
    \end{minipage}%
    \hfill
    \begin{minipage}{0.45\textwidth}
        \centering
        \includegraphics[width=\linewidth]{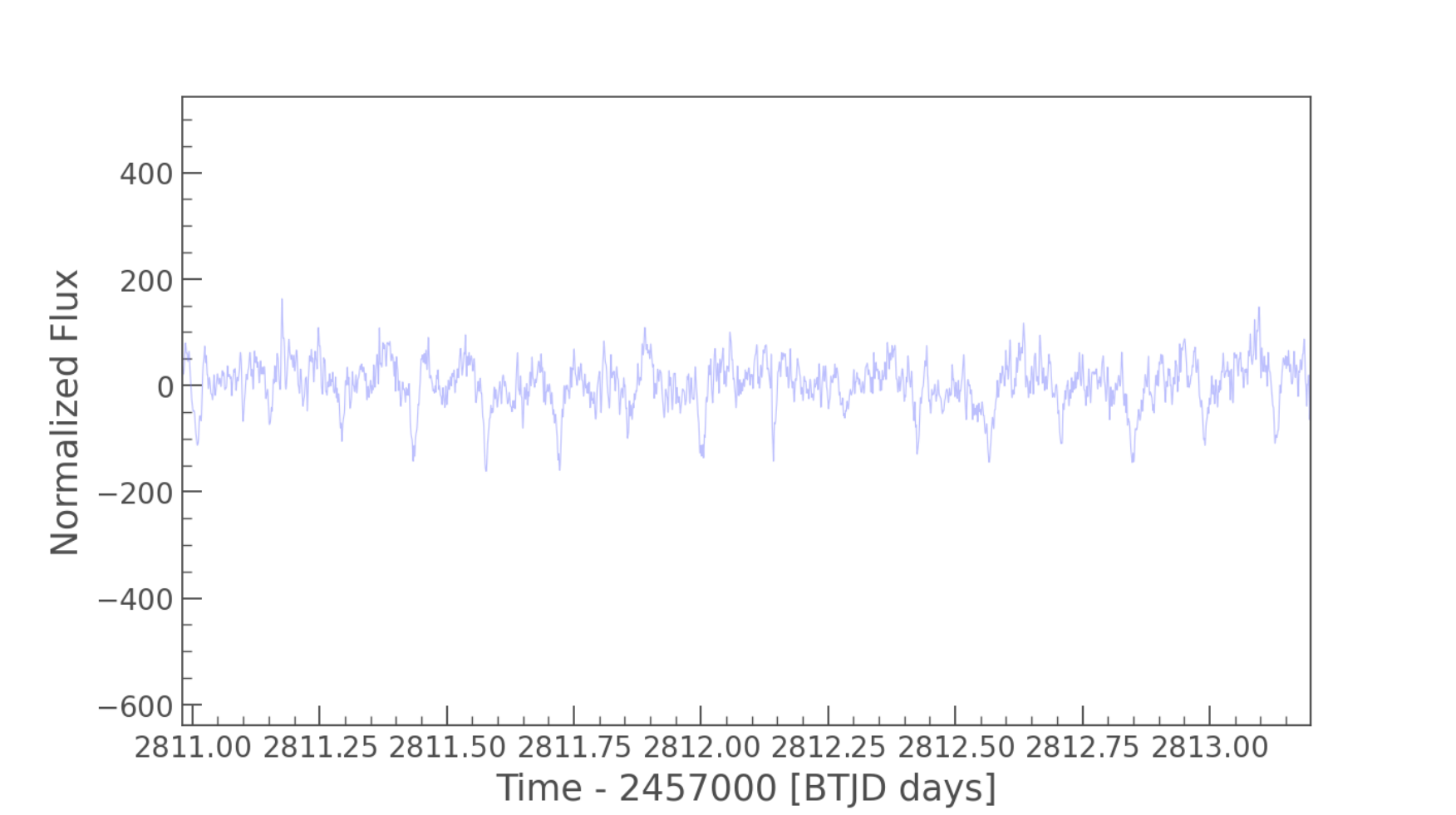}
        \caption{HBHA 4204--09}
    \end{minipage}
\caption{Detrended light curves for the six studied objects. Time is measured in Barycentric Julian Day (BTJD), and amplitude is expressed in normalized flux.}
\end{figure*}

    \label{fig:light_curves}

\clearpage

\subsection{Periodograms} 
\vspace{0.5em} 

\clearpage
\subsection{Sample Chunks} 
\vspace{0.5em} 

\clearpage
\subsection{O-C Diagrams} 
\vspace{0.5em} 

\end{document}